\documentclass[twocolumn]{aastex6}
\usepackage{xcolor}
\usepackage{txfonts,graphicx,rotating}
\usepackage{afterpage}

\usepackage{natbib,twoopt}

\begin{document}
\graphicspath{{./Figures/}}  

\DeclareRobustCommand{\ion}[2]{\textup{#1\,\textsc{\lowercase{#2}}}}

\newcommand{\sih}{\ensuremath{ {\rm [Si_{gas}/H}]}}
\newcommand{\HH}{\ensuremath{\rm H}}
\newcommand{\si}{\ensuremath{\rm Si}}
\def\sii{\ion{Si}{I}}
\def\siii{\ion{Si}{II}}
\def\ci{\ion{C}{I}}
\def\cii{\ion{C}{II}}
\newcommand{\aav}{\ensuremath{ \langle a \rangle_3} }
\newcommand{\mum}{\ensuremath{\rm \, \mu m}}
\newcommand{\Ms}{\ensuremath{\rm \,M_{\sun}}}
\newcommand{\Zs}{\ensuremath{\rm \,Z_{\sun}}}
\newcommand{\Mspc}{\ensuremath{\rm \,M_{\sun}\,pc^{-2}}}
\newcommand{\Msgyrpc}{\ensuremath{\rm \,M_{\sun}\,Gyr^{-1}\,pc^{-2}}}
\newcommand{\Msyr}{\ensuremath{\rm \,M_{\sun}\,yr^{-1}}}
\newcommand{\pyr}{\ensuremath{\rm \,yr^{-1}}}
\newcommand{\cmc}{\ensuremath{\rm \,cm^{-3}}}
\newcommand{\gcmc}{\ensuremath{\rm g\,cm^{-3}}}
\newcommand{\kms}{\ensuremath{\rm \,km\,s^{-1}}}
\newcommand{\ddt}[1]{{{\rm d}\, {#1} \over{\rm d}\,t}}

\author{Svitlana~Zhukovska\altaffilmark{1}, Clare~Dobbs\altaffilmark{2}, Edward B. Jenkins\altaffilmark{3}, Ralf~Klessen\altaffilmark{4}}
   \altaffiltext{1}{Max-Planck-Institut f\"ur Astrophysik, Karl-Schwarzschild-Str. 1, D-85741 Garching, Germany}
   \altaffiltext{2}{University of Exeter, Stocker Road, Exeter EX4 4QL, United Kingdom}
   \altaffiltext{3}{Princeton University Observatory, Princeton, NJ 08544-1001, USA}
   \altaffiltext{4}{Zentrum f\"ur Astronomie der Universit\"at Heidelberg, Institut f\"ur Theoretische Astrophysik, Albert-Ueberle-Str. 2, D-69120 Heidelberg, Germany}

\shortauthors{Zhukovska et al.}

\title{Modelling dust evolution in galaxies with a multiphase, inhomogeneous ISM}
 \shorttitle{Modelling dust evolution in a multiphase, inhomogeneous ISM}

  \begin{abstract}
We develop a  model of dust evolution in a multiphase, inhomogeneous ISM including dust growth and destruction processes. The physical conditions for grain evolution are taken from hydrodynamical simulations of giant molecular clouds in a Milky Way-like spiral galaxy. We improve the treatment of dust growth by accretion in the ISM to investigate the role of the temperature-dependent sticking coefficient and ion-grain interactions. From detailed observational data on the gas-phase Si abundances \sih{} measured in the local Galaxy, we derive a relation between the average \sih{} and the local gas density $n(\HH)$ which we use as a critical constraint for the models. This relation requires a sticking coefficient that decreases with the gas temperature. The relation predicted by the models reproduces the slope of $-0.5$ of the observed relation in cold clouds. It is steeper than that for the warm medium and is explained by the dust growth. We find that growth occurs in the cold medium for all adopted values of the minimum grain size $a_{\min}$ from 1 to 5~nm. For the classical cut-off of $a_{\min}=5$~nm, the Coulomb repulsion results in slower accretion and higher \sih{} than the observed values. For  $a_{\min}\lesssim 3$~nm, the Coulomb interactions enhance the growth rate, steepen the slope of \sih{}--$n(\HH)$ relation and provide a better match to observations. The rates of dust re-formation in the ISM by far exceed the rates of dust production by stellar sources. After the initial 140~Myr, the cycle of matter in and out of dust reaches a steady state, in which the dust growth balances the destruction on a similar timescale of 350~Myr.
  \end{abstract}


\maketitle

\section{Introduction}


A fraction of metals in the interstellar medium (ISM) of galaxies in both the local and high-redshift Universe resides in tiny refractory solid particles or dust grains \citep{Dorschner:1995p7228}. Interstellar dust constitutes less than 1\% of the ISM by mass, but it has manifold impact on the physics and chemistry of the ISM. Because dust locks some elements away from the gas phase, it reduces abundances of important gas coolants such as C$^+$  and Si$^+$ \citep{Bekki:2015hn, McKinnon:2016ft}. One of the most important roles of interstellar grains is that they facilitate the formation of molecular hydrogen (H$_2$) on their surfaces \citep{Hollenbach:1971hu}. The H$_2$ molecule is the main component of molecular clouds, which are the cradle of star formation in most of the Universe \citep{Klessen:2016ik}. Dust absorbs ultraviolet emission from young massive stars and re-emits it in the infrared, therefore the spectral energy distribution from dust is one of the primary indicators of star formation \citep{2013seg..book..419C}. Moreover, continuum emission from interstellar dust is a commonly used tracer of cold gas in galaxies \citep[e.g.,][]{Santini:2014fb}.

Despite the utmost importance of interstellar dust for the ISM evolution, high-resolution numerical simulations of the ISM  do not yet follow dust evolution. It is commonly assumed that dust abundance is scaled with metallicity and the scaling factor, grain size distribution and chemical composition have the average characteristics of dust in the local Galaxy \citep[e.g.][and references therein]{Walch:2015fg}. Galactic-scale chemodynamical models consider the dependence of galactic evolution on dust by including H$_2$ molecule formation on grain surfaces, but they follow the evolution of metallicity only and assume a fixed dust-to-metals ratio \citep[e.g.][and references therein]{Christensen:2012dw, Forbes:2016bu}.
 \cite{Hu:2016ek} recently investigated how the choice of dust-to-metal ratio value affects the evolution of dwarf galaxies through photoelectric heating process, keeping other dust properties fixed.

There is a strong observational evidence that cosmic dust properties are not universal.   \cite{RemyRuyer:2014uu} showed that there is large scatter in the observed dust-to-gas ratio vs. metallicity relation in local galaxies. Moreover, metal-poor dwarf galaxies tend to have lower dust-to-metal ratio compared to normal spiral galaxies. Recently, IR and far-IR surveys of dust emission opened a new venue to probe the spatial variations of dust properties on the small scale. \cite{RomanDuval:2014gu} analysed the IR emission maps of the Magellanic Clouds combined with various gas surveys and found large variations in the gas-to-dust ratio that are strongly correlated with the dust surface density distribution. Recently the far-IR emission survey with the Planck Satellite discovered similar variations of the gas-to-dust ratio both from cloud to cloud and within regions of individual clouds in the Milky Way \citep{Reach:2015gx}. An estimate of the true values of the dust-to-gas ratio at high surface densities is a challenging task, since different effects may cause variations of the apparent dust-to-gas ratio measured from observations: possible underestimation of the molecular gas mass because of the presence of CO-dark molecular gas, variations of dust emissivity in far-IR due caused by grain coagulation in dense clouds, and actual increase of the dust-to-gas ratio due to gas-grain interactions \citep[see][for discussion]{RomanDuval:2014gu}. Variations of dust properties from one line of sight to another in the local galaxy are also supported by observations of extinction curves \citep{Fitzpatrick:2007p6352}, dust opacities \citep{Roy:2013hm},  spectral characteristics \citep{Dartois:2004p466}, and interstellar element gas-phase abundances \citep{Savage:1996p486, Jenkins:2009p2144}. Thus, not only the dust-to-metal ratio, but also the composition and the size distribution of interstellar grains can differ from the standard. It is thus crucial to consider dust evolution in the context of modelling physics and chemistry of the ISM and using dust as a tracer to study the interstellar gas.

Evolution of interstellar dust is incorporated in one-zone models of the chemical evolution of galaxies, which consider dust properties averaged over the entire galaxy or over the vertical direction within concentric rings in spiral galaxies \citep[][]{Dwek:1980p490, Dwek:1998p67, Hirashita:1999p2082,  Zhukovska:2008bw, Zhukovska:2009p7232, Zhukovska:2013vg, Calura:2008p1752, Mattsson:2012kt, Asano:2013hg, deBennassuti:2014by, Hirashita:2015et}. These models demonstrate that the galactic dust content is mainly determined by the balance between dust destruction in interstellar shocks and dust production by evolved stars and growth by accretion of gas-phase species in the ISM. Given the relatively short timescales of dust destruction in the ISM of a few $10^8$~yr  \citep{Draine:1979p1036, Seab:1983gfa, Jones:1994p1037, Jones:1996p6593, SerraDiazCano:2008p588, Bocchio:2014es}, the dust growth in the ISM becomes the dominant dust source in galaxies on the timescale from a few $100$~Myr to a few Gyr \citep{Zhukovska:2014ey}. An accurate treatment of dust growth is therefore important for modelling of interstellar grains. Although simple dust evolution models can describe the average dust properties in these present-day galaxies, they do not consider the complex structure of the ISM and its impact on dust growth rate and are not able to describe the observed variations of dust properties across the ISM phases.

Recently, there have been a few attempts to take in to account the evolution of dust in numerical hydrodynamic galaxy simulations. \cite{Bekki:2013iw} included the time evolution of interstellar dust abundances in the chemodynamical model of disk galaxies and made a step forward by coupling it to the galactic evolution via H$_2$ formation on grain surfaces. The hydrodynamics of interstellar gas is modelled by the Smoothed Particle Hydrodynamics (SPH) method. With assumption that dust is coupled to gas, the model considers  variations of the dust-to-metal fraction in each gas particle. A disadvantage of the model of \cite{Bekki:2013iw} is the current implementation of dust processing in the ISM in which growth and destruction processes do not depend on the local conditions and occur on constant timescales. In following works, the dependence of the growth timescale on local conditions is included by scaling it with temperature and density of the particles \citep{Yozin:2014tt, Bekki:2015iy}. A new live dust particle model has been presented by \cite{Bekki:2015hn}, who decoupled gas and dust particles and implemented additional gas-grain interactions and radiation pressure on grains. Recently, dust evolution has been incorporated in the moving-mesh simulation code AREPO and used in zoom-in cosmological simulations of Milky Way-sized galaxies \citep{McKinnon:2016ft}. Their assumptions for the dust model are similar to those outlined in \cite{Bekki:2015iy}, with the exception of the destruction timescale, which they related to the local SN rate. While their simulations agree with a number of observables, they over-predict the dust-to-metal ratio in the circumgalactic medium. 

The hydrodynamic galactic simulations with evolving dust clearly demonstrate that grains influence evolution of galaxies, albeit not as strong as SN or active galactic nuclei feedback. However, the existing simulations should be improved in several ways. For example, they do not treat properly the dependence of the growth timecale on metallicity. The timescale of dust growth in the ISM is inversely proportional to the metallicity, resulting in the existence of the critical metallicity for the dust growth \citep[e.g.,][]{Zhukovska:2008p7215, Zhukovska:2009p7232, Asano:2013kl}. While some simulations include a dependence of the growth timescale on the local temperature and density \citep{Bekki:2015iy, Bekki:2015hn, McKinnon:2016ft}, they assume a fixed sticking coefficient over the entire simulation volume. With this assumption, grains can grow by accretion even in the warm or hot gas, If they stay sufficiently long time there ($\sim$1~Gyr).  However, sticking of the impinging species to the grain surfaces in this case is unlikely because of their high thermal energies \citep{DHendecourt:1985p867}. Moreover, the resolution of cosmological simulations ($\sim 10^5\Ms$) is not sufficient to investigate dense regions where dust growth is expected or to address the observed variations in dust abundances across the ISM phases in the local Galaxy.

In the current study, we describe a new model of dust evolution in the inhomogeneous ISM including dust destruction by SN shocks and dust growth in the ISM. We apply the model to study the evolution of the three-dimensional (3D) dust distribution in the local Galaxy using histories of physical conditions from  hydrodynamic simulations of the lifecycle of giant molecular clouds (GMCs) \citep{Dobbs:2013hb}. These simulations have sufficiently high resolution compared to the previous works to investigate the changes experienced by grains as they cycle between molecular clouds and ambient ISM. We substantially improve the treatment of dust growth in the ISM in two ways: by including a temperature dependent sticking coefficient and Coulomb interactions in calculations of the growth timescale. An advantage of our post-processing approach is that we can run multiple models to investigate how different model assumptions affect the resulting distribution of dust.  Our main goal is to use the analysis of the large amount of element depletion data measured in the local Galaxy \citep{Jenkins:2009p2144} to constrain the uncertainties in microphysics of the growth process.


The paper is organised as follows. We describe the initial conditions and main assumptions of these simulations in Sect.~\ref{sec:Simul}. Histories of physical conditions of gas parcels from the hydrodynamic simulations provide input for the dust evolution model. Section~\ref{sec:DustEvol} presents the formulation of the dust evolution model and our choice of model parameters. Section~\ref{sec:Depl} introduces the data on element depletions in the local Milky Way providing observational constraints for the model. Results of the model calculations of dust evolution as traced by Si element depletions are presented in Sect.~\ref{sec:Results}, where we discuss the timescales of dust destruction and formation, the rates of dust growth in the ISM as a function of ambient density and the distribution of the element depletion values. Sect.~\ref{sec:Depl-nComp} compares the theoretical and observed trends of Si depletion with density. Our conclusions are presented in Sect.~\ref{sec:Concl}.

\section{Simulations of lifecycle of giant molecular clouds}\label{sec:Simul}
In this paper, we utilise a hydrodynamic simulation of the ISM in a spiral Milky Way-like galaxy described in \cite{Dobbs:2013hb}. The simulation is performed using the SPH code sphNG \citep{Benz:1990ct, Bate:1995wm, Price:2007jo}. The gas in the simulation is subject to a galactic potential. We use a logarithmic galactic potential \citep{Binney:1987vb} which provides a flat rotation curve, and a two armed spiral potential from \cite{Cox:2002fz}. The spiral potential is fixed, the spiral arms rotating with a constant pattern speed of $19\, \rm km\, s^{-1}\, kpc$, and represents a perturbation of a few per cent compared to the logarithmic potential. GMCs form within the arms through a mixture of gravitational and thermal instabilities in the gas, and cloud-cloud collisions \citep{Dobbs:2008ez}. There is a total of 8 million gas particles, and each one has a mass $m_{\rm SPH}=312.5\Ms$. The gas is situated within a radius of $10\,$kpc, and the gas settles in approximate vertical equilibrium after $100\, $Myr (see \citealt{Dobbs:2011gp}). The average surface density of the gas is 8\Mspc. 

Cooling and heating of the ISM are incorporated according to \cite{Glover:2007p2460}, with temperatures representative of a multiphase medium, spanning a range $50\,$K to slightly above $10^6\,$K. The cooling processes include collisional cooling (mainly from C$^+$, O and Si$^+$, but H$_2$ cooling is also included) and the main heating process is photoelectric heating. The cooling is metallicity dependent, with the simulation assuming solar metallicity. Although in reality the metallicity would vary with radius, solar metallicity is consistent with the metallicity used in the dust post-processing which reflects the properties of the ISM at the solar radius. Self gravity of the gas is also taken into account.

 Stellar feedback is included in the form of supernovae feedback: once gas reaches a density of 500\cmc{} and is gravitationally collapsing. Thermal and kinetic energy are added to particles within a radius $R_s$ of about 15 pc from the densest particle, which is based on the smoothing length at these densities. The energy is inserted according to a snowplough solution, for the exact equations see the appendix of \cite{Dobbs:2011gp}. The amount of energy added is calculated as
\begin{equation}
     E_{\rm SN}=\frac{ \epsilon\, m(\HH)} {160} 10^{51} \, \rm erg,
\label{eq:mfeed}
\end{equation}
where $\epsilon$ is the star formation efficiency, chosen to be 0.05, $m(\HH)$ is the mass of molecular hydrogen within $R_s$, $10^{51}\,$erg is the energy of 1 supernova. We adopt a Salpeter IMF such that 1 supernova occurs for every 160\Ms{} of stars formed.

The hydrodynamic simulation then provides input for the dust evolution model such as the density, temperature and masses of the SPH particles at each stored snapshot  with cadence of 1~Myr. The particles which have undergone a recent supernova feedback event can also be identified. In particular, for each supernova event, the dust evolution model uses the total mass of gas where feedback is injected, 
 	\begin{equation}
		m_{j, \rm feed} = N_{j, \rm part} m_{\rm SPH},
	\end{equation}
where $j$ refers to the $j$th feedback event, $N_{j, \rm part}$ is the number of particles in the $j$th feedback event (typically around 20). All other assumptions about how the dust evolves in the ISM, and how it is destroyed by feedback, are applied in a post process step according to the dust evolution model described in the next Section.

\section{Dust evolution model}\label{sec:DustEvol}
In evolved metal-rich systems such as the present-day Milky Way, the timescales of enrichment of the ISM with chemical elements from stars are significantly longer than the timescales of mass transfer between the ISM phases. This allows us to make a number of simplifying assumptions and focus on the dominant sinks and sources of dust. Firstly, we fix the total (dust+gas) element abundances, and, secondly, we consider the dust production by growth in the ISM and neglect dust input from stars \citep{ODonnell:1997p683}. Additionally, we only include destruction of dust in the interstellar shocks and neglect destruction by star formation since its timescale of $2.5\,$Gyr is much longer than the current estimates for the dust lifetimes against destruction in shocks. We check the validity of these two latter assumptions by comparing the corresponding dust formation rates in Sect.~\ref{sec:Results}. 


We model the evolution of dust grains in the ISM by solving numerical differential equations for the degree of condensation, i.e. the fraction $f_X$ of the key element $X$ in dust. The key element is usually a constituent of the considered dust species which determines the reaction rate for dust growth. It is usually the species with the lowest gas-phase abundance. The $f_X$ is related to the logarithmic depletion as $\rm{[X/H]_{gas}} = \log(1-f_{\rm X}).$
The degree of condensation can be converted to the dust mass in a gas parcel as 	
	\begin{equation}
	m_{X, \rm dust} = \frac{A_{\rm X} \epsilon_X X_{\rm H} m_{\rm SPH}} {\xi_X} \cdot f_{X}, 
	\label{eq:m2f}
	\end{equation}
where $\epsilon_X$ and $A_X$ are the key element abundance and its atomic weight, and $\xi_X$ is the mass fraction of the key element in the considered solid, $m_{\rm SPH}$ is the mass of an SPH particle, and $X_{\rm H}=0.7$ is the hydrogen mass fraction in the gas.

Dust evolution is governed by growth and destruction processes, so that the change of the degree of condensation is
\begin{equation}
    \ddt{f_X}  = \left(\ddt{f_X}\right)_{\rm gr} + \left(\ddt{f_X}\right)_{\rm dest}.
\end{equation}
%
%

\begin{table}
\caption{Basic data used in model calculations}
\begin{tabular}{lc }
\hline\hline
Physical quantity & Value  \\
\hline
Key element & Si \\
$A_{\si}$ & 28 \\
$\eta_{\si}$    & 0.165 \\
Solid material density $\rho_{\rm d}$ (\gcmc) & 3.13 \\
Element abundance $\epsilon_{\si}$ ($10^{-5}$) & $3.24$ \\
Initial depletion \sih & $-0.5$ \\ 
Threshold for destruction in diffuse gas $n_{\rm dest}^{\rm diff}$ (\cmc) & 1\\
Timescale for destruction in diffuse gas $\tau_{\rm dest}^{\rm diff}$ (Gyr) & 0.1\\
\hline
\end{tabular}
\label{tab:PhysicalQuant}
\end{table}

\subsection{Growth in the ISM}
The mechanism of grain growth in the ISM is not fully understood. It has been suggested that accretion of gas-phase species on silicate and carbonaceous grains occurs selectively to keep them as two distinct populations \citep[see discussion in][]{Draine:2009p6616}. We therefore model the evolution of silicate dust independently of carbon dust. Note that, in the present work, we investigate the accretion of refractory elements on the grains which proceeds prior to the accretion of complex ice mantles and coagulation in dense cores of molecular clouds  \citep[e.g.][]{Joseph:1986p6009, Weingartner:1999p6573, Voshchinnikov:2010p6003}. 

One-zone dust evolution models usually assume that the dust growth occurs in molecular clouds based on their higher densities and consequently shorter collision timescales of gas-phase species with grain surfaces compared to the diffuse medium \citep{Dwek:1998p67, Zhukovska:2008bw}. We relax this assumption to include possible dust growth in the cold neutral medium (CNM) defined accordingly to the gas temperature range of 30 -- 300~K \citep[e.g.][]{Mihalas:1981vw}.

The change of the degree of condensation owing to the dust growth in the ISM by collisions of gas phase species with the grain surfaces is \citep{Zhukovska:2008bw}
	\begin{equation}
	 \left(\ddt{f_X}\right)_{\rm gr}  = \frac{1}{\tau_{\rm gr}} f_X(1-f_X).
	 \label{eq:dfdtgr}
	\end{equation}
Here $\tau_{{\rm gr},X}$ is the growth timescale of a given dust species
	\begin{equation}
	\label{eq:taugr}
		\tau^{-1}_{{\rm gr},X} = \frac{3 \alpha_X \epsilon_X m_X }{\rho_{\rm d} \xi_X \aav } \upsilon_{{\rm th},X} n_\mathrm{H},
	\end{equation}
where $\alpha_X$ is the sticking coefficient that expresses the probability of sticking of the growth species in collision with the surface, $\rho_{\rm d}$ is the density of the solid. The thermal velocity $\upsilon_{{\rm th},X}$ and the number density of the gas  $n_H$ are determined by the local conditions recorded in the particle histories. The average grain radius is
\begin{equation}
 \aav = \langle a^3 \rangle/ \langle a^2 \rangle,
\label{eq:a3normal}
\end{equation}
where 
\begin{equation}
\langle a^l \rangle \sim \int {{\rm d}  n_{\rm gr}(a)} \big/  { {\rm d}a}\, a^l {\rm d}a,
\end{equation}
 is the $l$th moment of the grain size distribution $ {{\rm d}  n_{\rm gr}(a)} \big/  { {\rm d}a}$ defined so that $ {\rm d}  n_{\rm gr}(a) $ is the number of grains with radii from $a$ to $a+{\rm d}a$.

In the diffuse ISM, most of the key species are singly ionised, therefore electrostatic interactions can enhance or reduce  accretion rates, depending on the grain charge. To include this effect of Coulomb interactions in our model, we adopt the enhancement factors $D(a)$ from work of \cite{Weingartner:1999p6573} calculated for the standard interstellar UV radiation field, collisions with singly ionsed gas species and the grain charge distribution described in \cite{Weingartner:2001p461}. 
We include enhancement effects in the CNM using a modified average grain radius in Eq.~(\ref{eq:taugr})
\begin{equation}
	\langle a \rangle_3 = \langle a^3 \rangle/ \int  \frac{{\rm d} n_{\rm gr}(a)}{{\rm d}a} a^2 D(a) {\rm d}a,
	\label{eq:a3}
\end{equation}
where the electrostatic factor $D(a)$ accounts for the change in the cross section of an interaction between ion and grain. For neutral particles in MCs, $D(a)=1$.

The timescale of dust re-formation in the ISM is defined as
\begin{equation}
\tau_{\rm form} =  \frac{M_{\rm dust}}{\dot{M}_{\rm dust, gr}}, 
\end{equation}
which for the simulations with equal mass particles can be written using summations over all particles
\begin{equation}
    \tau_{X, \rm form} =  \sum\limits_{i=1}^{N_{\rm tot}} f_{X,i}  \,\Big/ \,  \sum\limits_{i=1}^{N_{\rm tot}} \left( \ddt{f_{X,i}}\right)_{\rm gr}.
    \label{eq:taudest}
\end{equation}
We assume that dust mass grows in the quiescent ISM and exclude the gas particles which undergo SN feedback event described in the next section.

\subsection{Dust destruction in SN shocks}\label{sec:ModDustDestr}
Sputtering of grains in high velocity shocks ($\gtrsim 100\ \kms$) is the main mechanism of dust destruction in the ISM. The large scale simulations used in this work do not resolve the scale on which gas is shocked to high velocities, therefore our approach to calculations of dust destruction utilises the existing SN energy feedback scheme  described in Sect.~\ref{sec:Simul}. 

In each feedback event, the energy from SNe is equally distributed between the gas  particles with the total mass $m_{j, \rm feed}$ given by Eq.~(\ref{eq:mfeed}). In the same feedback event, dust is completely destroyed in a mass of gas $m_{\rm cleared}$ which for the homogeneous ambient medium is defined as  \citep{McKee:1989p1030}
	\begin{equation}
	 m_{\rm cleared}(n_0) = \int_{\upsilon_0}^{\upsilon_f} \varepsilon(\upsilon_s,n_0) 
	 \left| \frac{dM_s(\upsilon_s, n_0)}{d\upsilon_s} \right| d\upsilon_s ,
	\label{eq:MassCleared}
	\end{equation}
where $\upsilon_0$ and $\upsilon_f$ are the initial and final velocities of the SNR expanding into an ambient medium of density $n_0$, respectively, $\left| \frac{dM_s(\upsilon_s, n_0)}{d \upsilon_s} \right| d \upsilon_s$ is the mass of gas swept up by a shock with velocity in the range of $[\upsilon_s,\upsilon_s+ d \upsilon_s]$, $\varepsilon$ is the degree of dust destruction in a gas parcel shocked to the velocity $\upsilon_s$. To calculate $m_{\rm cleared}$ for each SN remnant, we use Eq.~(\ref{eq:MassCleared}) together with an analytical expression for $M_s(\upsilon_s,n_0)$ given by \cite{Dwek:2007p496} which describes the adiabatic expansion and pressure-driven snow-plough stages of the SNR evolution.

With the resolution of our hydrodynamic simulations, $m_{\rm cleared}$ is lower than $m_{j, \rm feed}$. We therefore make two assumptions which allow us to simplify the treatment of dust destruction. We approximate the total dust mass destroyed in the $j$th event by $m_{\rm cleared}$ and assume the dust content is reduced equally in all affected gas particles by a fraction
	\begin{equation}
		\eta_{j, \rm dest} = \frac{m_{\rm cleared}}{m_{j, \rm feed}}.
	\end{equation}
The change in the condensation fraction in each gas particle in the $j$th feedback event is $-\eta_{j, \rm dest} f_X$. The change of condensation degree in a gas parcel which experienced the $j$th feedback event is   
	\begin{equation}
		 \left(\ddt{f_X}\right)_{\rm dest}^{\rm feed} = -\frac{\eta_{j, \rm dest} f_{X}}{\Delta t},
		 \label{eq:DestrRateFeed}
     \end{equation}	
where ${\Delta t}$ is the time step following the SN explosion. By using Eq.~(\ref{eq:m2f}) and summing over all particles affected by the total $N_{\rm feed}$ feedback events in GMCs, we derive the total dust mass destruction rate 
	\begin{equation}
		 \left(\ddt{M_{X,\rm dust}}\right)_{\rm dest}^{\rm feed} = - \frac{A_{\rm X} \epsilon_X X_{\rm H} m_{\rm SPH}} {\xi_X} \sum_{i=1}^{N_\mathrm{feed}} \sum_{j=1}^{N_{j, \rm part}}  \frac{\eta_{j, \rm dest} f_{X,i,j}}{\Delta t_{i,j}} ,
	\label{eq:DestrRateFeedtot}
	\end{equation}
where $N_\mathrm{feed}$ is the total number of feedback events that occurred between two subsequent snapshots.

The current implementation of stellar feedback in our simulations neglects such forms of stellar feedback as stellar winds and radiation, prostellar jets and SNe exploding in low-density regions. The impact of different mechanisms of stellar feedback on the dynamics of the ISM is discussed by \cite{Walch:2015fg}. Among different processes,  blast waves in the low-density gas are most important for dust destruction \citep{McKee:1989p1030, Draine:1990p495}. Observationally it has been determined that 20--25\% of OB stars belong to the field stars and will likely explode in the diffuse ISM \citep{Oey:2012ts}.  Without account of these SNe we may underestimate the overall dust destruction in the ISM. To investigate their effect on the outcome of simulations, we also run models with an additional term in the total destruction rate for all particles with density below 
$n_{\rm dest}^{\rm diff}=1\cmc$ adopted as an upper limit for the diffuse ISM
	\begin{equation}
	\label{eq:dfdtdiff}
		 \left(\ddt{f_{X}}\right)_\mathrm{dest}^\mathrm{diff} = -\frac{f_{X}}{\tau_\mathrm{dest}^\mathrm{diff}},
	\end{equation}
where $\tau_\mathrm{dest}^\mathrm{diff}$ is the timescale to  destroy all dust in the diffuse medium. This destruction rate is added on each time step only to those particles which are not affected by current stellar feedback events in GMCs. Note that in order to derive the timescale to destroy all grains in the ISM by this process $\tau_\mathrm{dest}^\mathrm{diff}$ has to be divided by the mass fraction of diffuse gas.


\begin{figure}
\includegraphics[width=0.5\textwidth]{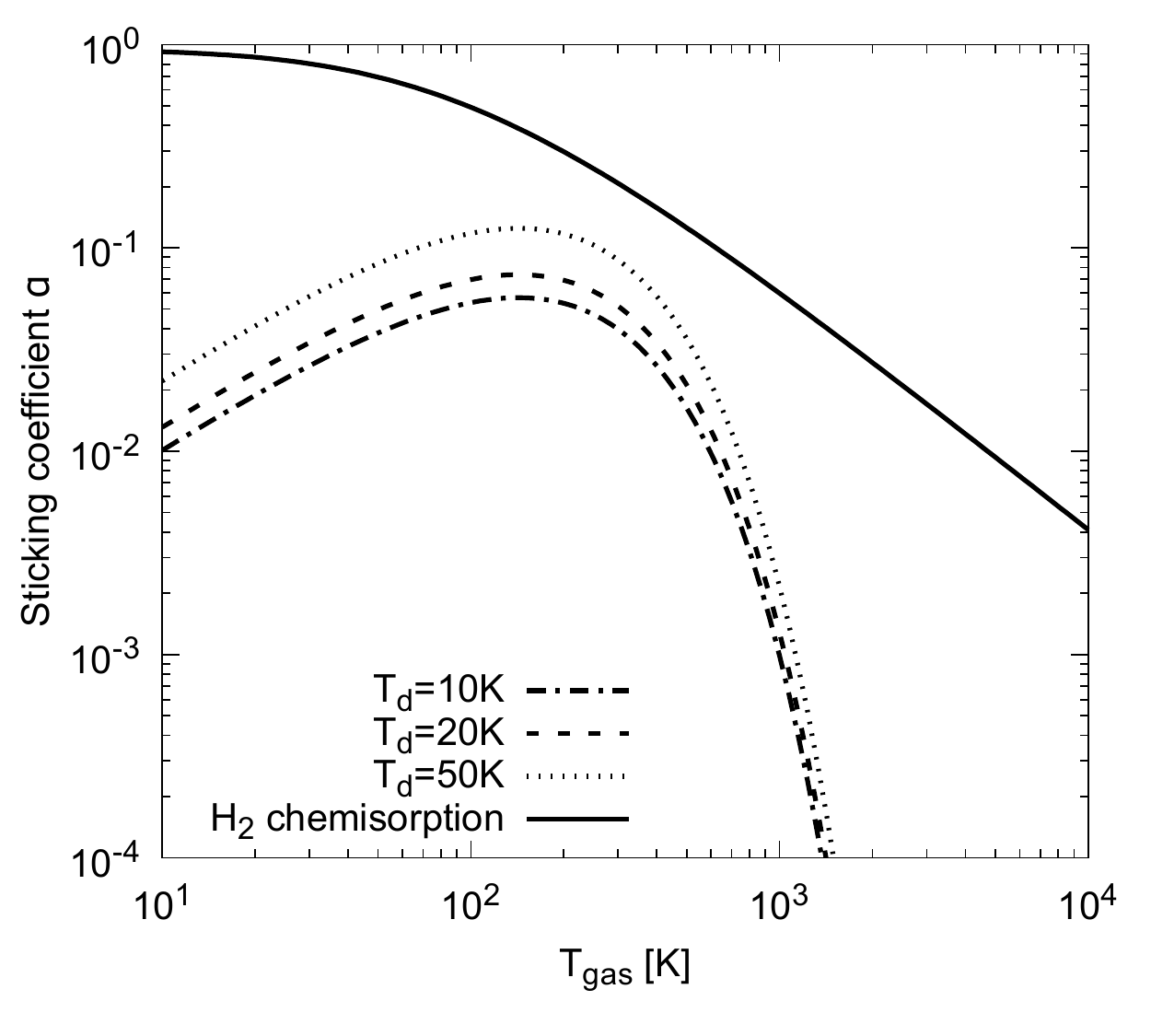}
\caption{Sticking coefficient for H$_2$ chemisorption on the silicate surface derived by \cite{Chaabouni:2012gw} (solid line) and that for physisorption derived by \cite{LeitchDevlin:1985tp} for dust temperatures 10, 20 and 50~K.}
\label{fig:StickCoef}
\end{figure}


\begin{table}
\begin{center}
\caption{Average grain radius \aav adopted in model calculations for the selected minimum sizes of the MRN grain size distribution}
\begin{tabular}{lccc }
\hline\hline
 $a_{\min}$ (nm) &  \multicolumn{2}{c}{\aav (nm) } \\
  & normal\tablenotemark{a} & enhanced\tablenotemark{b}  \\
\hline
 1 & 15.8 & 0.8 \\
 3 & 27.4 & 7.7  \\
 5 &  35.4 & 48.2 \\ 
\hline
\end{tabular}
\label{tab:avsizes}
\end{center}
\tablenotetext{a}{The average grain radius derived from Eq.~(\ref{eq:a3normal}).}
\tablenotetext{b}{\aav derived from Eq.~(\ref{eq:a3}) for the enhancement factor $D(a)$ for silicate grains in the CNM.} 
\end{table}

\subsection{Model parameters}\label{sec:ModelParam}
To solve equations for the condensation degree numerically, we need to make assumptions about the main properties of dust. For the total (gas+dust) chemical composition, we adopt solar metallicity $\Zs=0.015$ and element abundance ratios from \cite{2009LanB...4B..712L}.  In this work, we focus on growth of silicate dust, assuming that Si is the key species in silicate growth.  For the adopted olivine-type composition (MgFeSiO$_4$) of silicates, the mass fraction of Si in dust is $\eta_{\si}=0.165$. All physical quantities used in the calculations are summarised in Table~\ref{tab:PhysicalQuant}.

For the initial dust abundance, we take a constant value for all gas particles $f_{\rm Si}=0.68$ ($\sih=-0.5$). It is on the lower limit of the general depletion level in the WNM derived by \cite{Savage:1996p484} and somewhat lower than the value $\sih=-0.36$ suggested by the depletion data from work by \cite{Jenkins:2009p2144} (see Sect.~\ref{sec:Depl}). This value is consistent with the Si abundance required by models for extinction, emission, and polarisation of light by dust in the diffuse ISM in the solar neighbourhood \citep{Siebenmorgen:2014ga}. We intentionally take low initial dust abundances for all particles with higher density, to explore how the system reaches the balance between destruction and production processes. 

Efficiencies of grain destruction $\varepsilon$ for different dust species are obtained from extensive numerical calculations of grain evolution in the shocked gas. For  $\varepsilon(\upsilon)$, we adopt the values from works by \cite{Jones:1994p1037} and \cite{Jones:1996p6593} derived for steady shock models. 

The dust growth rate depends on the choice of the grain size distribution through Eq.~(\ref{eq:taugr}). For simplicity, we apply a commonly used power law distribution ${\rm d} n_{\rm gr}(a)/{\rm d}a \sim a^{-3.5}$  \citep[][, hereafter called MRN]{Mathis:1977p750}, which runs from $a_{\min}=5$~nm to $a_{\max}=250$~nm. The value $a_{\min}=3$~nm is used in our reference models. To investigate how the minimum grain size affects our results, we perform additional model calculations for $a_{\min}=1$~nm and 5~nm and discuss them in Sect.~\ref{sec:ResMinSize}. The corresponding values of the average grain size are listed in Table~\ref{tab:avsizes}.

\subsubsection{Description of the models}
The basic name of our models is "MRN$x$nm'', where $x$ is the minimum grain size in nm and MRN indicates that all models assume MRN grain size distribution. 
We add "E" at the beginning of the name of models in which the average timescale of accretion \aav is calculated with account of electrostatic interactions due to ion-grain collisions in the CNM, otherwise a fixed value of  \aav is used in all phases (Table~\ref{tab:avsizes}). 
The models with additional destruction in the diffuse medium as described in Sect.~\ref{sec:ModDustDestr} are denoted with an additional prefix "C". The model  ECMRN3nm is our best fit model for which most of the analysis is given below.

\subsubsection{Sticking coefficient}\label{sec:Modalpha}
The sticking coefficient $\alpha$ is the major source of uncertainty of our model. One-zone dust evolution models usually assume that growth occurs in molecular clouds and key species colliding with the grain surface stick perfectly, so that $\alpha=1$ (\citealt{ODonnell:1997p683, Zhukovska:2008bw, Asano:2013hg}; see also \citealt{Weingartner:1999p6573}). \cite{Hirashita:2011jr} adopt a more conservative value of $\alpha=0.3$. A constant value of $\alpha$ has been also adopted in recent numerical simulations of galaxies with dust evolution \citep[e.g.,][]{Bekki:2015hn, McKinnon:2016ft}. This assumption is justified at low gas temperature $T_{\rm gas}\approx 10-100$~K and dust temperatures  $T_{\rm dust}\approx 10$~K  because the kinetic energy of  incident Si atoms is significantly lower than their binding energy on the surface \citep[e.g.,][]{DHendecourt:1985p867}. It may however overestimate the dust growth rate in the warm medium in hydrodynamical simulations. 

In this work, we include gas-grain interactions in the diffuse ISM where most of the silicon is singly ionised. The sticking coefficient for Si$^+$ on silicate surface is not known. \cite{Watson:6p6307} discuss possible outcomes of the interaction between a positive ion and a negatively charged grain. They point that when an ion approaches the grain surface, one of the electrons in the grain may tunnel through the work function and neutralise the ion before it reaches the surface. The probability that the atom will remain on the surface upon collision increases dramatically if it can be chemisorbed \citep{Watson:6p6307}. This possibility is supported by recent experiments in Jena which demonstrate that adsorbed species can form chemical bonds characteristic for silica and silicates at substrate temperatures of 10~K without an activation barrier \citep{Krasnokutski:2014bi, Rouille:2014gi}. 

The sticking coefficient depends on many parameters including poorly understood surface properties of interstellar grains. With these uncertainties, the observed relation between interstellar depletions and gas density may provide an observational constraint for the choice of $\alpha$. To this end, we investigate various choices of the sticking coefficient: a) a fixed value $\alpha=1$ for $T_{\rm gas} \leqslant  300$~K and $\alpha=0$ for $T_{\rm gas}>300$~K; b) a fixed value $\alpha=1$ for all temperatures; c) experimentally measured $\alpha$ for chemisorption of H$_2$ molecules on silicate surface by \cite{Chaabouni:2012gw}; and d) $\alpha(T_{\rm gas}, T_{\rm dust})$ from theoretical calculations of physisorption performed by \cite{LeitchDevlin:1985tp}. In absence of estimates of $\alpha$  for physisorption of \si{} on silicate grain surfaces, we provisionally adopt the data for carbon atoms arriving on graphite lattice from work by  \cite{LeitchDevlin:1985tp}, in the functional form derived by \cite{Grassi:2011gg}. The dust temperature is fixed to 20~K.

We deem the temperature dependence of the sticking coefficient measured by \cite{Chaabouni:2012gw} more plausible than that of the $\alpha$ derived by \cite{LeitchDevlin:1985tp} (Fig.~\ref{fig:StickCoef}). The former increases at low gas temperatures and approaches 1 at $T_{\rm gas} \lesssim 10$~K, while the latter peaks at $T_{\rm gas} \approx 200~K$ and decreases at lower gas temperatures. However, after initial testing and comparison of model predictions with the observational constraints discussed in Sect.~\ref{sec:Depl}, we assume the case a) with a simple step function dependence of $\alpha$ on $T_{\rm gas}$  for the reference model. It restricts the growth with temperature and does not depend on the exact shape of $\alpha(T_{\rm gas})$ that has yet to be determined by future experiments.

As will be shown later in the manuscript the assumption of $\alpha=1$ for all temperatures results in too high average depletions compared to the range of values derived from depletion studies and should not be used. 

\begin{figure}
\includegraphics[width=.5\textwidth, page=1]{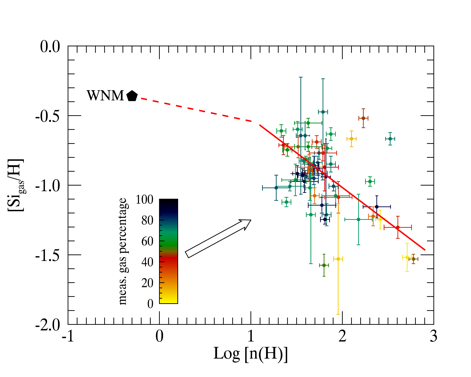}
\caption{{\it Multiple points with error bars:\/} Gas-phase \si{} abundances ${\rm [\si_{gas}/\HH]}$ \citep{ Jenkins:2009p2144} versus the corresponding logarithms of the local gas densities $n(\HH)$, as inferred from absorption features arising from the \ci{} fine-structure levels in various directions \citep{Jenkins:2011by}. These measurements are indicated with colors that correspond to estimates for the fractions of the total material that were sampled by the fine-structure population ratios.  A solid red line indicates the least-squares linear fit to the data.  {\it Single black pentagon:\/} An estimate for the representative \si{} abundance for the warm neutral medium (WNM) for a value of $F_*$ equal to 0.12 that corresponds to depletions measured by  \cite{Savage:1996p486} for this phase.} 
\label{fig:Depl-nh-obs}
\end{figure}

\section{Depletion of elements in the ISM}\label{sec:Depl}
In this Section, we discuss observational data used to constrain our model. Interstellar dust abundances can be studied by analysing the gas-phase abundances of refractory elements, assuming that the elements missing from the gas are locked in dust.  The abundances of free atoms and ions in space can be determined by analysing absorption features appearing in the ultraviolet spectra of background stars \citep{Spitzer:1975ca}. The logarithmic depletion of an element $X$ in the ISM is the gas-phase element abundance relative to a reference abundance, for which we adopt the element abundance in proto-Sun reflected by the present-day photospheric abundance with an increase of 0.07~dex to account for gravitational settling \citep{Lodders:2003bf}
\begin{equation}
{\rm \left[ \frac{X_{gas}}{H} \right] } = \log\left(\frac{n_{\rm X}}{n_H}\right)_{\rm gas} - \log\left(\frac{n_{\rm X}}{n_H} \right)_{\sun}.
\end{equation} 

There are significant variations of interstellar element depletions between different lines of sight, which are probably related to dust evolution driven by local conditions and the recent history of the gas. The observed determinations of element depletions \sih{} can therefore be used to validate and provide useful constraints for our model. 

One simple quantity commonly used in the interpretation of interstellar element depletions is the average gas density $ \langle n(\HH) \rangle$ along the sight line, defined as the ratio of the column density of hydrogen (both atomic and molecular) to the distance of the background star $D$
\begin{equation}
 \langle n(\HH) \rangle =[ N_{\HH} + N_{\rm H_2}]/D
\end{equation}

This is a volume-weighted determination of local densities.  It has been well established that $ \langle n(\HH) \rangle$ strongly correlates with element depletions \citep[e.g.,][]{Joseph:1986p6009, Joseph:12p7198, Jenkins:1987wz, Savage:1996p484, Jenkins:2009p2144, Haris:2016bg}. However, this correlation may be governed more by the relative proportions of some representative high density material compared to very low density gas \citep{Spitzer:1985hr, Jenkins:1986cd}, rather than an average for the local densities over regions containing mostly neutral hydrogen along the sight line. In reality, the value of $\langle n(\HH) \rangle$ provides a lower limit for the gas densities associated with measured element depletions, with an understanding that the true average hydrogen density for the measurements may be significantly higher if much of the sight line is filled with fully ionised (hence invisible) gas.

The outputs from our dust formation models show details on the distributions of \si{} depletions as a function of local density.  The lack of an exact correspondence between $\langle n(\HH) \rangle$ and representative local densities presents a challenge in our making meaningful comparisons of the observed depletions to the dust models.

To derive the relation between \sih{} and the mean sight-line density $\langle n(\HH) \rangle$, we start with the data compiled by \cite{Jenkins:2009p2144} which included over 243 lines of sight probing a large range of physical conditions. The sight lines with extraordinary high velocities typical for shocked gas were excluded from the analysis. Only stars situated within about 1.5~kpc from the solar Galactocentric radius were used for quantitative analysis, which avoids the possible consequences of the Galactic radial abundance gradient. In the work of \cite{Jenkins:2009p2144}, the depletions were expressed in terms of a factor designated as $F_*$, a parameter that characterised the relative strengths of some available element depletions for a given line of sight.  Owing to the fact that all of the different element depletions behave in well defined manners with respect to $F_*$, this parameter can be used as a proxy to determine the depletion of any particular element (in our case \si{}), even if that element was not observed for the sight line under consideration.  We may express how $F_*$ relates to $\langle n(\HH) \rangle$ using a linear fit to the data plotted Fig.~16 in \cite{Jenkins:2009p2144} and the \si{} abundance \sih{} as a function of $F_*$ (by the use of Eq.~(10) in \cite{Jenkins:2009p2144}), which allows us to derive the following relation between the gas-phase \si{} abundance and the average density 
\begin{equation}
 \sih =-0.524 \log \langle n(\HH) \rangle - 1.10.
\label{eq:SiDepl_n09}
\end{equation}

There is a more direct way to probe the local gas density in diffuse cold clouds based on an analysis of the collisional excitation of the fine-structure levels of interstellar neutral carbon. \cite{Jenkins:2011by} investigated the \ci{} absorption lines seen in the spectra of stars recorded in the highest resolution echelle modes of the Space Telescope Imaging Spectrograph on board the Hubble Space Telescope.  This investigation included 89 out of the 243 lines of sight from the data sample from \cite{Jenkins:2009p2144}. Using the estimates of thermal gas pressure and temperatures evaluated by \cite{Jenkins:2011by}, we calculate the gas densities $n_{\HH}$ in the interstellar cold clouds and paired them with the gas-phase \si{} abundances along the same sight lines.  The outcomes are shown in Fig.~\ref{fig:Depl-nh-obs}.  The best fit to these results is represented by the equation
\begin{equation}
 \sih =-0.497 \log n(\HH) - 0.0225,
\label{eq:SiDepl_n11}
\end{equation}
and it is shown as a solid red line in the figure.  This outcome is based on minimizing the $\chi^2$ for both the errors in $\log n(\HH)$ and  \sih{} using the routine {\tt fitexy} \citep{Press:2007vx}.

As a result of the fact that the ionization equilibrium between \ci{} and \cii{} is governed by the local electron density, the measurements of the fine-structure level populations for \ci{} are biased in favor of denser gas, which means that the derived hydrogen densities are restricted to the range from about $10\cmc$ to $10^3\cmc$.  It follows that the WNM is not sampled by the \ci{} because the characteristic densities are much lower.  For this reason, we must try to understand empirically how influential this WNM contamination is for our determinations of representative local densities of neutral hydrogen that appear in Fig.~\ref{fig:Depl-nh-obs}.

For every sight line, \cite{Jenkins:2011by} estimated the fraction of gas that was sampled by the \ci{} fine-structure excitations by examining the absorption profiles of other species in their dominant ionization stages, which could serve as proxies for C II. Based on this information, they stated this sampling fraction in terms of a quantity that they called "fraction of C II observed"(see their Table 3). If this percentage is low, one must conclude that the WNM dominates over the CNM, and this effect could lead to an outcome for the silicon depletion that is less severe than that for the dense region \citep{Savage:1996p486}.

To evaluate the importance of the WNM contributions, we re-evaluated the best-fit relationship only for cases where the fraction of \cii{} in the measurable  \ci{} velocity range exceeded 50\%.  The change of the coefficients in Eq.~(\ref{eq:SiDepl_n11}) was inconsequential. 

In their review of interstellar abundances, \cite{Savage:1996p486} stated general values for depletions in the WNM in the Galactic disk, which Jenkins converted to a value for the $F_*$ parameter equal to 0.12, which in turn yields $\sih=-0.36$. We include in Fig.~\ref{fig:Depl-nh-obs} a point that shows this level of Si depletion at a location for a typical density of the WNM, $n(H) \approx 0.5 \cmc$. The value $n(\HH)\sim 0.5\,\cmc$ for the representative density is justified by \cite{Jenkins:2013kb} -- see Section 7.4 of that paper. Gas in the Galactic halo probably represents material that has been shocked and accelerated, and here the \si{} abundances may be even higher because the grains have been eroded even further.  

\begin{figure}[t]
\includegraphics[width=.5\textwidth, page=2]{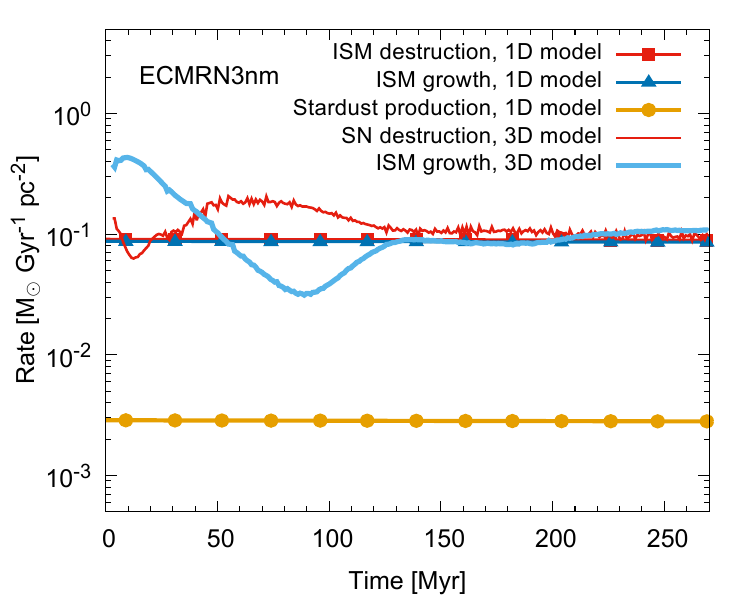}
\includegraphics[width=.5\textwidth, page=2]{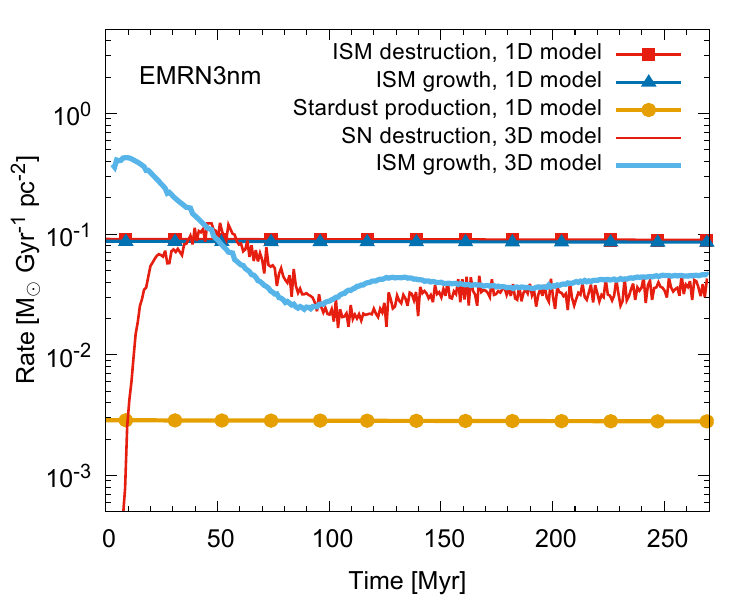}
\caption{Evolution of rates of dust destruction in the interstellar shocks and production by growth in the ISM (red thin and blue thick lines, respectively). The rates of dust production in the ISM, destruction by SN shocks and stardust injection rate from 1D dust evolution model \citep{Zhukovska:2008bw} are shown for comparison (lines with filled triangles, squares and circles, respectively). Top and bottom panels show models with and without additional destruction in the diffuse ISM (ECMRN3nm and EMRN3nm, respectively).}
\label{fig:rates}
\end{figure}

\begin{figure}
   \includegraphics[width=.5\textwidth, page=1]{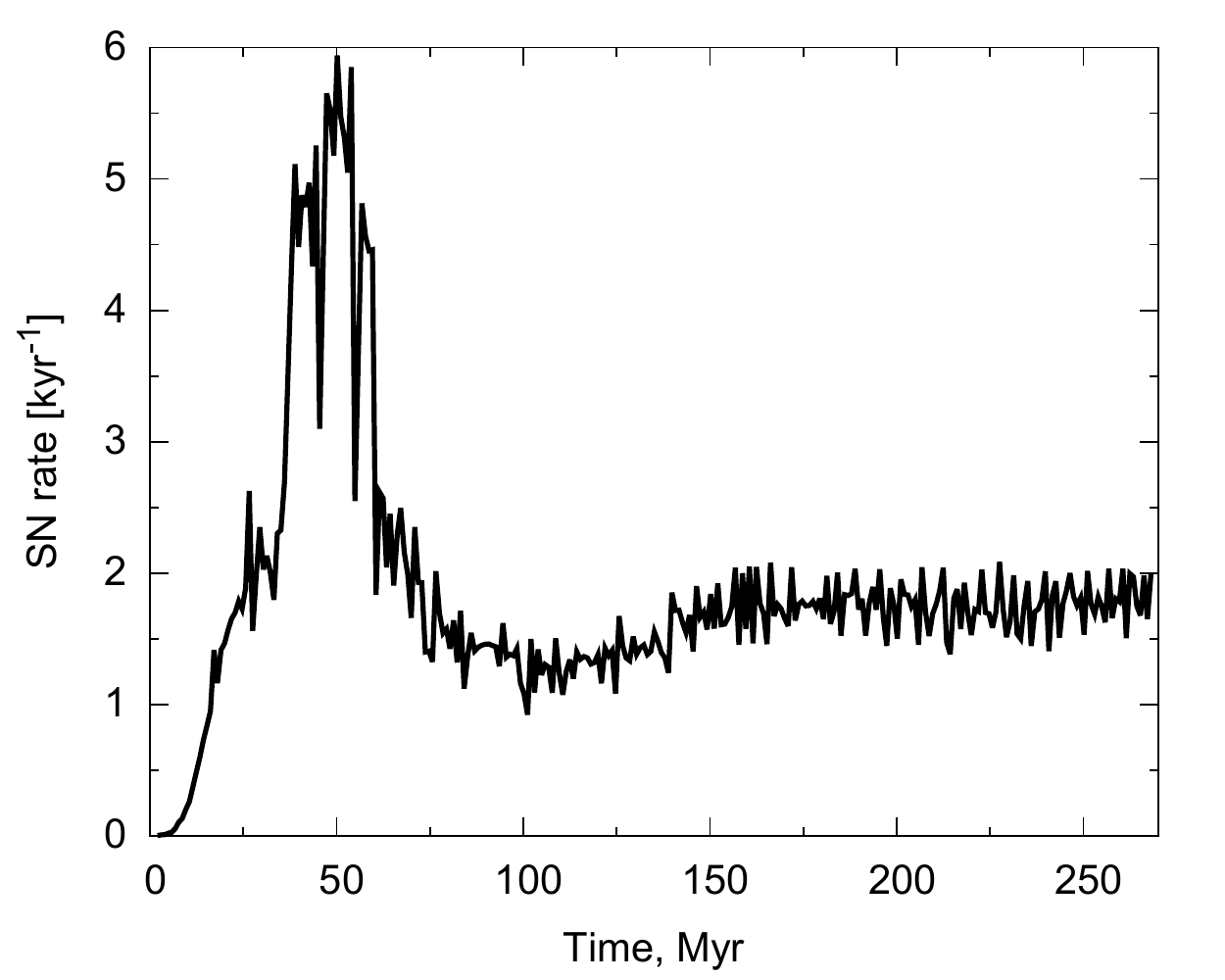}
    \caption{Evolution of the total supernova rate in the simulation volume.}
    \label{fig:snrate}
\end{figure}

\section{Results}\label{sec:Results}
In the following, we present the results of the calculations of dust evolution as traced by Si abundance in the Milky-Way-like disk galaxy simulated for 270~Myr. As our goal is to compare the model predictions with the observed depletion data limited to a few kpc around the Sun \citep{Jenkins:2009p2144}, we select the particles in a ring with galactic radii from 6 to 9~kpc representing the conditions similar to those in the solar neighbourhood.

\subsection{Dust production rates}\label{sec:DPR}
Our model predicts the rates of destruction and production of silicate dust with a fixed olivine-type composition. If we assume that the carbon-to-silicate dust mass ratio does not vary significantly across the ISM phases, we can estimate the total rates of dust production and destruction in the ISM. We adopt the carbon-to-silicate ratio of 0.5 inferred for interstellar dust in the local diffuse ISM \citep{Dwek:2005p1018}. 

Figure~\ref{fig:rates} shows the variations of the interstellar dust production and destruction rates per unit of the surface area as a function of time after the start of our calculations. On two panels, we show the results for models EMRN3nm and ECMRN3nm corresponding to the case of dust destruction only by feedback from massive stars in GMCs and the case with the additional destruction in the diffuse medium.

The total dust growth rate is higher at the beginning because of the low values of the initial element depletions in the simulation volume. The initial peak is followed by the dip in the production rate at 90~Myr resulting from the enhanced SN rate (Fig.~\ref{fig:snrate}) caused by the initial conditions of the hydrodynamic simulations. 

Figure~\ref{fig:rates} illustrates that, after the initial 140~Myr, the cycle of matter in and out of dust reaches a steady state, in which interstellar dust is distributed over the ISM phases in such a way that the dust production rate balances the destruction rate. The destruction and production rates converge to a similar value in the steady state controlled by the destruction timescale. It is about $0.04\Msgyrpc$ for models EMRN3nm, which determines a lower limit for the dust destruction rate and corresponds to destruction only by feedback from massive stars in GMCs. For ECMRN3nm model, the value of destruction/production rates in the steady state is by a factor of 2.5 higher and attains the value of  $0.1\Msgyrpc$.

The resulting total rate of dust destruction/ISM growth for ECMRN3nm model matches remarkably well the value from a simple one-zone model of chemical evolution of the solar neighbourhood with dust derived by \cite{Zhukovska:2008bw}. The one-zone model considers the evolution of dust surface densities averaged over the vertical direction as well as in a 1~kpc wide ring with the solar galactocentric radius. It follows the evolution of dust and gas chemical abundances for 13~Gyr, starting from the primordial chemical abundances to the present day values. 
Time variations of the destruction and ISM dust growth rates predicted by the one-zone model are also displayed in Fig.~\ref{fig:rates}. For comparison with our results, we take the rates from the one-zone model for the last 270~Myr before the present time. These rates from the one-zone model do not noticeably vary over the considered time span because it is much shorter than the present-day timescale of chemical enrichment. 

1D dust evolution model also predicts the dust injection rate by AGB stars and SNe based on the star formation history of the the solar neighbourhood and stellar dust yields described in \cite{Zhukovska:2008bw} and \cite{Ferrarotti:2006p993}. The current rate of dust input from stars is $3\times 10^{-3} \Msgyrpc$ (Fig.~\ref{fig:rates}). This is 33 times lower than the dust production rate by ISM growth in our reference model ECMRN3nm. This corroborates our model assumption that growth in the ISM is the dominant dust source in the present-day Milky Way and dust production in stellar winds can be neglected.  Note that this assumption does not hold during the early evolution of the Milky Way or in young metal-poor dwarf galaxies \citep{Zhukovska:2008bw, Zhukovska:2014ey}.

\begin{figure}[t]
\includegraphics[width=.5\textwidth, page=3]{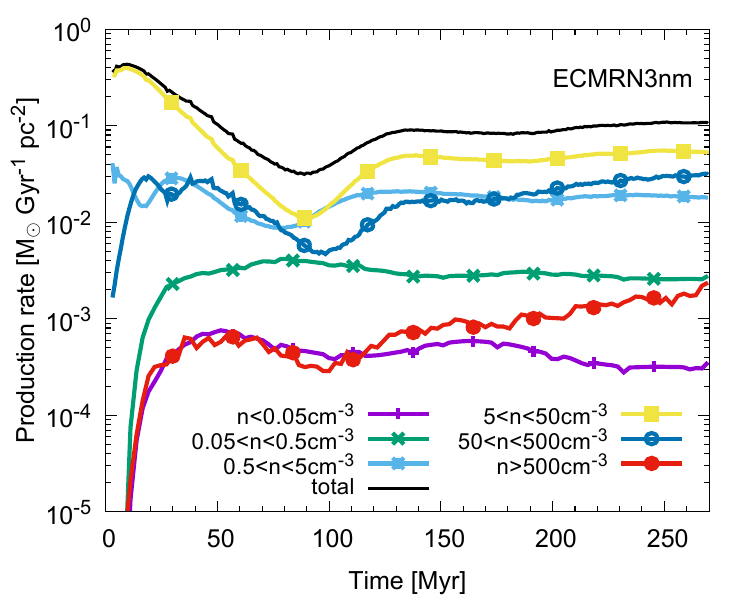}
\caption{Evolution of dust production rates in the ISM in model ECMRN3nm separated into gas density bins of width of 1 dex from 0.05 to 500\cmc.}
\label{fig:Ratesn}
\end{figure}

\begin{figure}[t]
     \includegraphics[width=.5\textwidth, page=1]{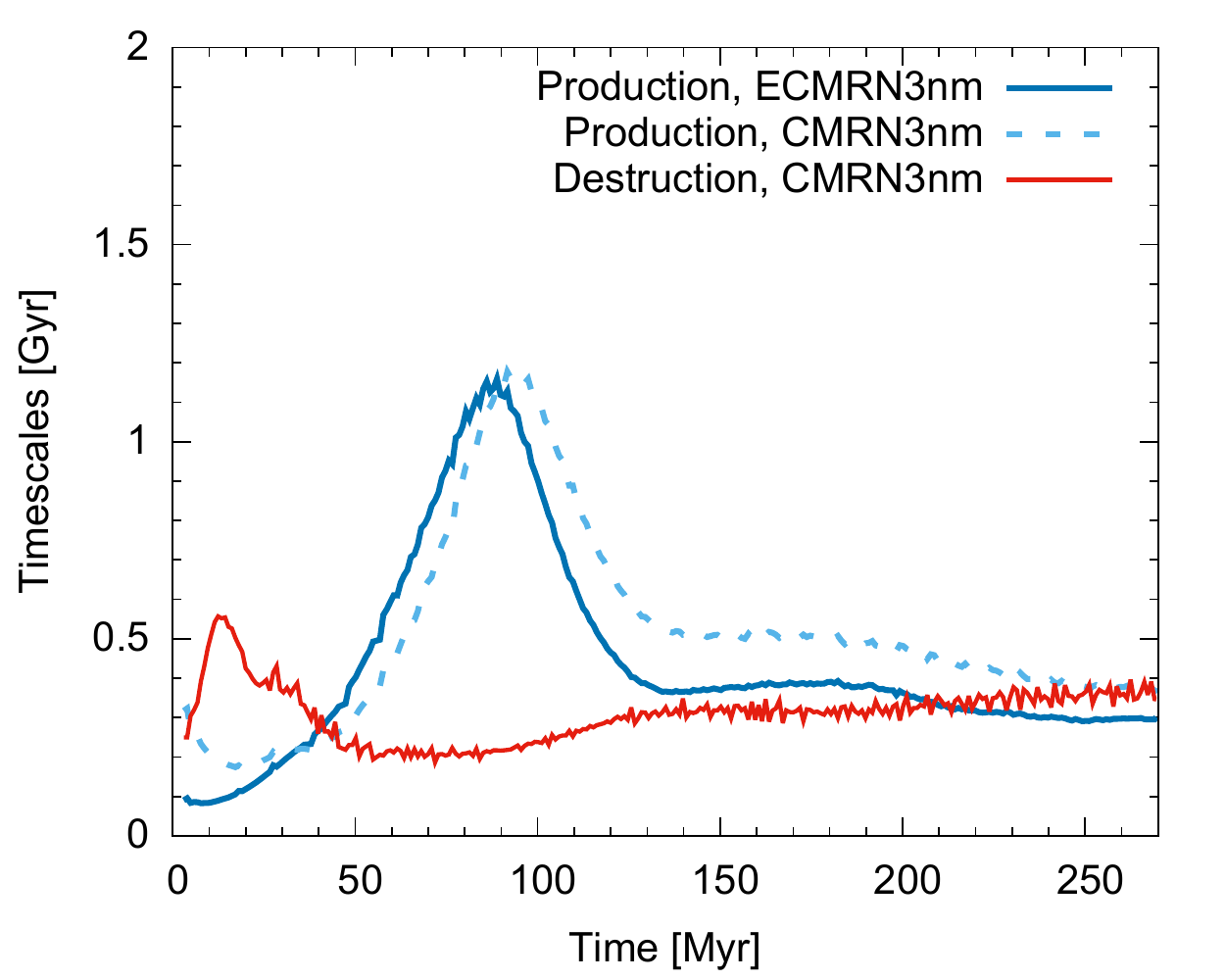}
     \includegraphics[width=.5\textwidth, page=1]{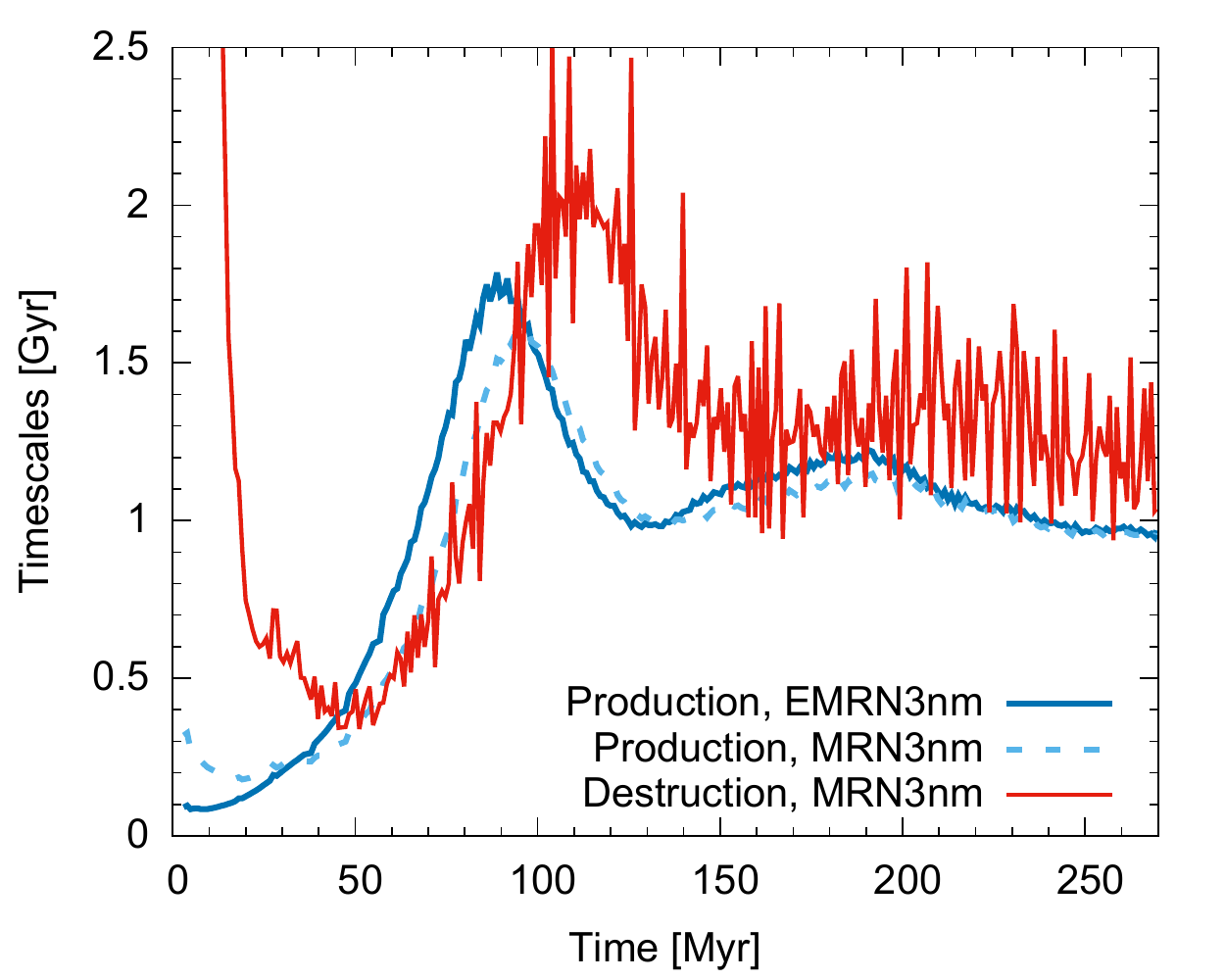}
     \caption{\textit{Top panel.} Evolution of timescales of silicate dust destruction and production by growth in the ISM for model ECMRN3nm (red thin and blue thick lines, respectively). Production rate for model CMRN3nm is shown with dashed line. \textit{Bottom panel.} The same for the EMRN3nm and MRN3nm models without additional destruction in the diffuse ISM.
     }
\label{fig:timesc}
\end{figure}


\subsubsection{Where do interstellar grains grow?}
To investigate the conditions that favour dust growth in our hydrodynamical simulations, we compare the dust production rates within 
the logarithmic gas density intervals  from 0.05\cmc{} to 500\cmc{} with a stepsize of 1~dex. Time evolution of the rates in each density bin for ECMRN3nm model is shown in Fig.~\ref{fig:Ratesn}. It demonstrates that most of growth occurs at densities $5\cmc < n \leqslant 50\cmc$, which is 49\% of the total ISM growth rate. Gas with densities  $50 \cmc < n \leqslant 500\cmc$ contributes 30\% and more diffuse clouds with $0.5 \cmc <n \leqslant 5\cmc$ 17\% to the total growth rate, correspondingly. The contribution from the diffuse medium with $n \leqslant 0.5\cmc$ constitutes only 3\% of the total rate. This is not surprising, since the growth is limited to $T \leqslant 300$~K in the reference model. 

Our results do not drastically change for models with different temperature-dependent sticking coefficient tested in this work  \citep{Chaabouni:2012gw, LeitchDevlin:1985tp}. Compared to a simple step function dependence on temperature in the reference models, the sticking coefficient usually decreases with  gas temperatures from the maximum value close to 1 similar to the behavior of the data from \cite{Chaabouni:2012gw} shown in Fig.~\ref{fig:StickCoef}. Such dependence somewhat reduces the growth rate in the density bin $0.5 \cmc <n \leqslant 5\cmc$ and increases it in the next density bin ($5\cmc < n \leqslant 50\cmc$) resulting in very similar values of growth rates in these two bins. 

 Our simulations do not resolve the dense gas well, therefore its actual  contribution can be higher than the present value of a few per cent.

\begin{figure}
\includegraphics[width=.5\textwidth, page=1]{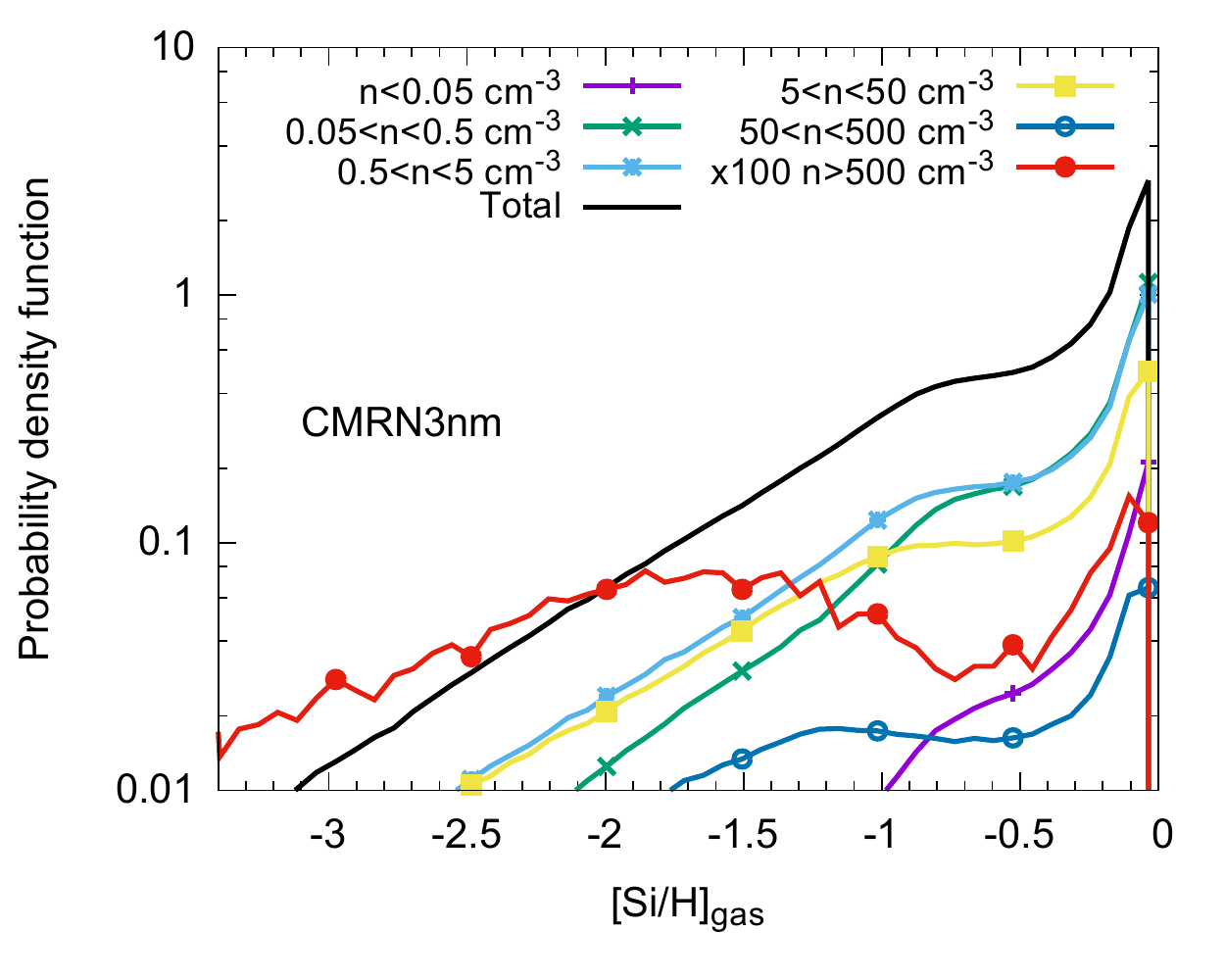}
\includegraphics[width=.5\textwidth, page=2]{PDFs_depl-MRN3nm-r69}
\caption{Probability density functions of the \si{} depletions per unit log depletion for the final simulation snapshot split in different gas density intervals spanning from 0.05 to 500\cmc (solid lines with various symbols, as indicated). Black line shows the PDF for all densities. All PDFs are normalised to the total gas mass in the considered volume. Top and bottom panels display models CMRN3nm and ECMRN3nm, respectively.}
\label{fig:PDFSidepl}
\end{figure}

\subsection{Timescales of dust destruction and re-formation in the ISM}\label{sec:ResTimesc}

Figure~\ref{fig:timesc} shows the evolution of the average timescales of silicate dust destruction and re-formation in the ISM derived using the dust production and destruction rates for models ECMRN3nm and EMRN3nm (Fig.~\ref{fig:rates}). The early evolution of the dust formation timescale is strongly affected by the initial conditions for both the hydrodynamical simulations and dust abundances. The adopted constant depletion $ \sih=-0.5$ at the onset of the simulations causes a rapid growth phase with the formation timescale of a few hundred Myr. Later on, the growth is slowed down as a consequence of a temporal reduction of the GMC mass fraction by feedback from the initial burst of star formation. This reduction in the star formation activity manifests in Fig.~\ref{fig:snrate} as a dip in the SN rate following the peak corresponding to the initial starburst. 
Strong feedback from the starburst suppresses the growth in clouds and increases $\tau_\mathrm{form}$ to over 1~Gyr by a time of 90~Myr in both models. However, the steady state values of the destruction and formation timescales, reached after the balance between disruption and formation of GMCs is established, differ significantly for the two models.

The average timescale of dust destruction  $\tau_\mathrm{dest}$ calculated for model ECMRN3nm is 3 times shorter than for model EMRN3nm and reaches down to 350~Myr. For the adopted low value of the timescale of the destruction rate in the diffuse medium $\tau^\mathrm{diff}_\mathrm{dest}=100$~Myr, $\tau_\mathrm{dest}$ is determined by the $\tau^\mathrm{diff}_\mathrm{dest}$ and the mass fraction of the diffuse ISM in our simulations which is about 20\%. Efficient destruction in the diffuse phase with a short $\tau^\mathrm{diff}_\mathrm{dest}$, resulting in the short lifetime of grains in the ISM, are required for our model to reproduce the mean \si{} depletion in the diffuse medium. Our results thus re-enforce the assumption proposed in earlier studies that sputtering of grains in the diffuse ISM by single SNe is the dominant mechanism of dust destruction \citep{McKee:1989p1030}. 

When only destruction by SN feedback in GMCs is included (model EMRN3nm), the lifetime of grains in the ISM $\tau_\mathrm{dest}$ at the end of simulations reaches 1~Gyr. Model EMRN3nm overpredicts the high depletions of Si in the WNM compared to the observed value, $-1$ and $-0.36$, respectively. The value of 350~Myr predicted by model ECMRN3nm at the end of simulations is 50~Myr lower that the lifetime of silicate grains in the Milky Way estimated by \cite{Jones:1996p6593}. Following \cite{McKee:1989p1030}, they combined the observed mass of the ISM and galactic SN rate with their estimate of the $m_\mathrm{cleared}$ given by Eq.~(\ref{eq:MassCleared}) resulting from extensive numerical calculations of the dust destruction in the blast wave.

We can also compare the timescale from our simulations with other studies which improved various aspects of the physics of grain destruction in a shock, but kept the same simple approach to the grain cycle in the homogeneous ISM. \cite{Bocchio:2014es}, for example, re-evaluated the dust lifetimes against destruction using models with a better treatment of the dust dynamics in the shock. With consideration of uncertainties in the observed mass of the ISM and SN rate, they derived the lifetime of $310\pm270$~Myr for silicate grains, which is very close to the value predicted by our best fit model. \cite{Slavin:2015in} proposed much longer lifetimes of silicate grains of 2--3~Gyr. Although they included the hydrodynamical shock evolution in calculations of the destruction efficiency $\varepsilon(\upsilon_s,n_0)$, the new value of $m_\mathrm{cleared}$ from their work is not very different from the previous works. The main reason for their longer $\tau_\mathrm{dest}$ is the new estimates of the total gas mass and SN rate from recent observations.


The average dust formation timescale in the ISM $\tau_{\rm form}$ for models with enhanced collision rates of cations with grains in the CNM is not very different from those without it (Fig~\ref{fig:timesc}). The differences are larger for models with  additional destruction in the diffuse medium. They can be explained by differences in $\tau_\mathrm{gr}$ and element depletion levels in these models. The growth timescale in the CNM $\tau_{\mathrm{gr}}$ is 4 times shorter for model ECMRN3nm which results in rapid depletion of \si{} from the gas phase to higher levels than in model CMRN3nm. Since $\tau_{\rm form}$ increases with $\tau_\mathrm{gr}$ in individual gas particles and decreases with their element condensation fractions $f_{\si}$, see Equations~(\ref{eq:dfdtgr}) and (\ref{eq:taugr}), the interplay between these two factors determines the ratio of $\tau_{\rm form}$ predicted by the models with and without enhanced collision rates.
 

\subsection{Distribution of Si depletion with density}\label{sec:PDFs}
We analyse the complex distribution of Si depletion values in the simulation  volume by means of mass-weighted probability density functions (PDFs) calculated in logarithmically equal gas density intervals. Figure~\ref{fig:PDFSidepl} displays the PDFs of the final distribution of Si depletions for the logarithmic gas density intervals  from 0.05\cmc{} to 500\cmc{} with steps of 1~dex. All PDFs are normalised to the total mass of gas. The main feature of the derived distributions is broad asymmetric profiles which become shallower in higher gas density bins as a consequence of the dust growth. Increase is commensurate with the observed depletion trends  discussed in detail in Sect.~\ref{sec:Depl-nComp}. 

The broadest depletion distribution derived for the densities $n>500\cmc$ reflects rich dynamic histories of particles at these densities. Particles with the highest depletions belong to clouds which were formed or evolved from the disruption of larger clouds. They have spent longer time at larger densities. On the other hand, the low density tail of the PDF corresponds to gas which has recently undergone compression from the diffuse gas so that the depletion values have not yet substantially increased. The highly asymmetric profiles of PDFs shown in Fig.~\ref{fig:PDFSidepl} with a steady increase towards the less severe depletions are a direct consequence of the additional dust destruction in the diffuse phase. The mean depletions for these PDFs nevertheless provide the best match to the observed depletion values.

For the selected minimum grain size of 3~nm, the enhanced collision rates due to electrostatic focusing modestly affects the PDFs of Si depletion (Fig.~\ref{fig:PDFSidepl}, bottom panel). This process is responsible for the shallower slopes of PDFs for the density range $0.05\cmc< n \leqslant 50\cmc$ owing to the enhanced growth under these conditions. The PDF for $n>50\cmc$ looks lower for the ECMRN3nm model, because a larger fraction of gas at these densities has  \si{} depletion values below $-3.5$.


\begin{figure}[t]
\includegraphics[width=.5\textwidth, page=1]{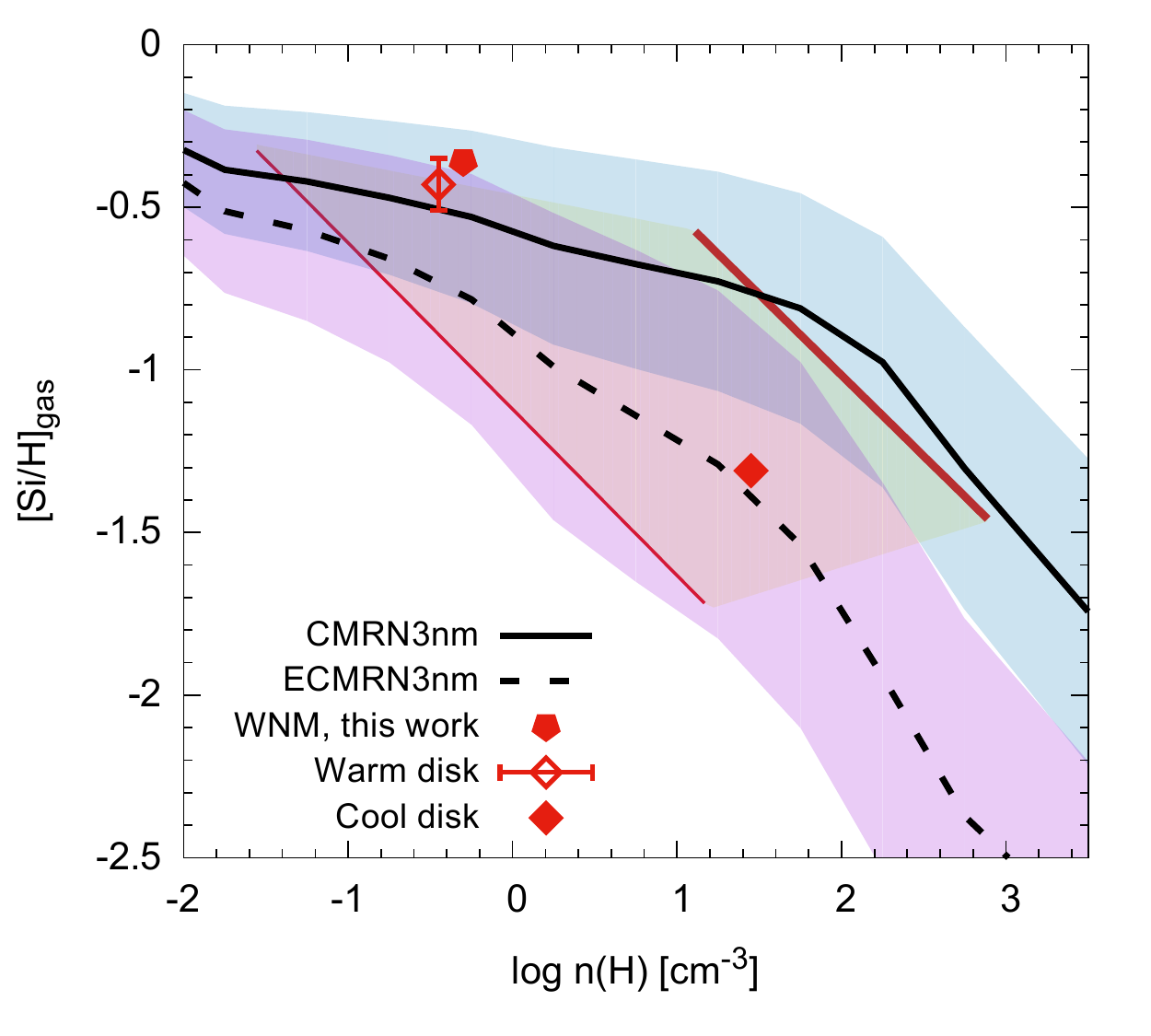} 
\caption{Relation between mean Si depletion and gas density derived using the final PDFs of gas-phase \si{} abundances for models CMRN3nm and ECMRN3nm (solid and dashed black lines, respectively). The filled areas around the synthetic relations indicate the corresponding standard deviations. The pentagon symbol indicates a generalised depletion level at a location for a typical density of the WNM from observations. The thick red line shows the linear fit to the observed data given by Eq.~(\ref{eq:SiDepl_n11}) and the thin red line shows the lower limit for this relation derived for the mean gas density on the line of sight given by Eq.~(\ref{eq:SiDepl_n09}) (see explanation in Sect.~\ref{sec:Depl}). The yellow area shows the range of values between two relations. Typical values for the warm and cool galactic disk (open and filled diamond, respectively) derived by \cite{Savage:1996p486} are shown for comparison.}
\label{fig:Depl-logn}
\end{figure}

\begin{figure}[t]
\includegraphics[width=.5\textwidth, page=2]{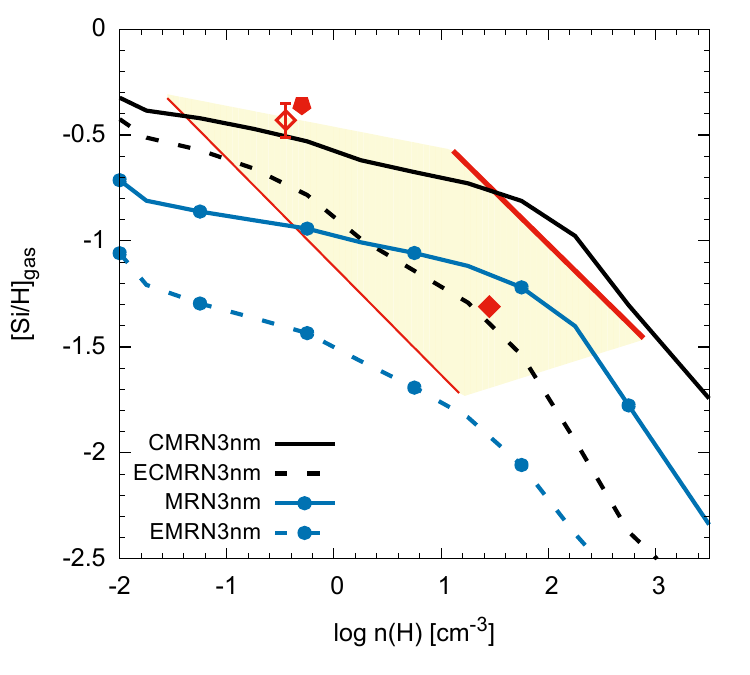} 
\caption{The same as in Fig.~\ref{fig:Depl-logn} compared to the relations between mean Si depletion and gas density derived for models MRN3nm and EMRN3nm (blue solid and dashed lines with circles, respectively) without additional dust destruction in the diffuse ISM. We do not show the standard deviations for the synthetic relations for clarity of the figure.}
\label{fig:Depl-logn-nodestr}
\end{figure}

\subsection{Si depletion -- gas density relation}\label{sec:Depl-nComp}
In order to compare our model predictions with the observed relation between Si depletion and gas density (Sect.~\ref{sec:Depl}), we compute the mean values and standard deviations of the  interstellar Si depletion using the PDFs for the final simulation snapshot. For this analysis, the PDFs are computed in the same way as described in the previous section, but within much finer logarithmic gas density intervals. The resulting relations between the gas phase Si abundance and gas density are shown in Fig.~\ref{fig:Depl-logn} for reference models ECMRN3nm and CMRN3nm, in which a simple step function is adopted for the sticking coefficient ($\alpha=1$ for $T_{\rm gas}<300$~K and  $\alpha=0$ for higher $T_{\rm gas}$).

Figure~\ref{fig:Depl-logn} also shows the relations between the mean \sih{} values and the gas density derived from observations. The relation based on the gas densities probed by \ci{} fine structure lines (Eq.~(\ref{eq:SiDepl_n09})) is limited to a narrow density range of $10-10^3\cmc$. A generalised Si depletion level in the WNM inferred from observations provides a constraint for dust evolution models in the WNM density regime. Additionally, Figure~\ref{fig:Depl-logn} shows  the relation derived for the mean gas density on the lines of sight given by Eq~(\ref{eq:SiDepl_n09}), which we consider as a lower limit to the observed relation. 

A conspicuous feature of the synthetic mean \sih--$n(\HH)$ relation  for all models is the large standard deviation of about 0.5 dex at the lowest gas densities up to one dex at higher densities. These large variations predicted by the dust evolution models agree well with the dispersion in the real \sih{} values from observations (Fig.~\ref{fig:Depl-nh-obs}). The dispersion arises from different dynamical histories of gas parcels with the same density that undergo various stages of the lifecycle of GMCs. Gas that previously resided in GMCs on its way to the WNM has higher Si depletions, while matter from the WNM has lower element depletions because of dust destruction process. This effect can be also seen in the widths of PDFs calculated within the larger density bins discussed above. 

Another characteristic feature of the synthetic mean \sih--$n(\HH)$ relation is its different slopes at the low and high gas densities seen as the double power law in Fig.~\ref{fig:Depl-logn}. The slope in the dense gas regime is steeper and its value of 0.5 is in excellent agreement with the slope of the observed relation. This agreement provides the evidence of dust growth by accretion in the ISM. With enhanced collision rates due to ion-grain interactions (model ECMRN3nm), the accretion timescale is shorter and the slope in the diffuse gas is therefore steeper compared to the CMRN3nm model. 

Comparison of the slopes at low densities for models ECMRn3nm and CMRN3nm reveals that this slope also depends on the accretion timescale, although the growth is limited to temperatures below 300~K in these models. This is also supported by the fact that dust destruction in the diffuse phase does not depend on the local density in our model. The slope at low density regime is determined by the dust depletion at high densities, which is controlled by the growth timescale. Additional important factor controlling the distribution of dust abundances at low densities is the mass circulation between the GMCs and diffuse medium. In this work, it is set by numerical simulations of GMC evolution and is the same in both dust models, hence the differences in the slope at low densities between models CMRN3nm and EMRN3nm as well as in other models discussed in the remainder of this paper are the "memory" of dust growth by accretion at higher densities.


In the reference models described above we include dust destruction in the diffuse ISM in addition to destruction by type II SNe in GMCs. We compare the synthetic relations between \sih{} and $n(\HH)$ obtained from the reference models with those from models MRN3nm and EMRN3nm in Fig.~\ref{fig:Depl-logn-nodestr}. All parameters are the same with the exception of dust destruction, which in models MRN3nm and EMRN3nm occurs only through feedback from massive stars in GMCs. Models without additional destruction in the diffuse ISM overpredict Si depletion by 0.5 dex in the case without ion-grain interactions included and by 1 dex in the model EMRN3nm with enhanced collision rates due to ion-grain interactions.  

The observed distribution of gas phase element abundances thus implies that some fraction of stellar feedback energy is injected in the diffuse medium. SN energy input only in the dense phase results in too low \sih{} values compared to the observations. The importance of stellar feedback energy injection in the diffuse ISM is also stressed in work by \cite{Walch:2015fg}, who analyse how location of SN explosions affects the structure of the ISM by means of high resolution hydrodynamical simulations of the ISM.

The present dust evolution model does not include dust and gas input from stars. Since the volume filling factor of the diffuse medium is much larger than that of the molecular clouds, evolved stars inject matter mainly in the diffuse phase. For stardust evolution model of \cite{Zhukovska:2008bw}, the dust input rate from stars  is significantly lower than the rate of dust mass growth in the ISM (see Sect.~\ref{sec:DPR}) and \sih{} in stellar ejecta is higher than the observed value in the warm phase. Therefore, adding stellar dust sources self-consistently would result in slightly higher values of  \sih{} in the diffuse gas, but would not change the slope of the \sih--$n(\HH)$ relation.

In this work, we focus on the matter cycle at the solar galactocentric radius. \cite{Dobbs:2013hb} demonstrate that the dominant mechanisms of disruption and formation of the  GMCs are different in the outer and inner disk. Clouds in the inner disk are more likely to get destroyed by stellar feedback and sheer, while large scale gas dynamics is more important on the larger radii. The radial distribution of dust and, correspondingly, element depletions in the gas phase are affected by the changes in dynamical evolution of GMCs with distance from the Galactic centre, in addition to the metallicity radial gradients. Our preliminary study indicates that shorter lifetimes of the clouds in the outer disk  result in lower element depletions. We investigate how the radial variations in the dynamical evolution of clouds impact the dust abundances and the \sih--$n(\HH)$ relation in a forthcoming paper.


\begin{figure}
\includegraphics[width=.5\textwidth, page=3]{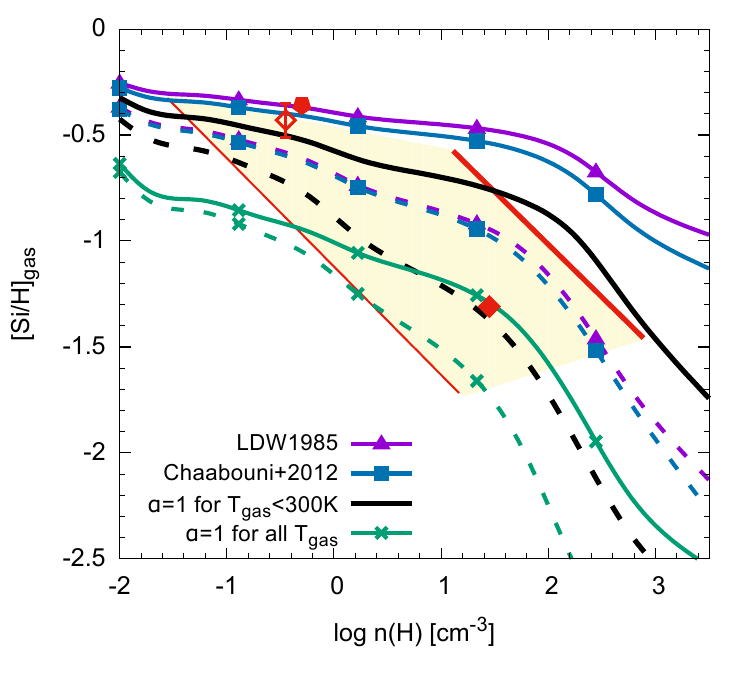}
\caption{Relation between Si depletion and gas density derived using the PDFs of gas-phase \si{} abundances for the final snapshot in the simulations for different sticking coefficients: $\alpha$ from work of \cite{LeitchDevlin:1985tp}, \cite{Chaabouni:2012gw} and $\alpha=1$ (lines with  triangles, squares and crosses, respectively). Solid and dashed lines show models CMRN3nm and ECMRN3nm, respectively. The observational data are displayed in the same way as in Fig.~\ref{fig:Depl-logn}.}
\label{fig:Depl-logn-alpha}
\end{figure}

\subsection{Observational constraints for the sticking coefficient}

The sticking coefficient affects the final distribution of the interstellar depletions by limiting the densities of the ISM at which dust mass can grow by accretion  of gas-phase species. In order to analyse how the modelled depletions depend on the choice of the sticking coefficient, we perform the simulations for models with the same parameters as ECMRN3nm and CMRN3nm, but assuming different $\alpha$ as discussed in Sect.~\ref{sec:Modalpha}. The resulting relations for the mean depletion as a function of gas density are shown in Fig.~\ref{fig:Depl-logn-alpha}.


The reference model includes the dust growth only for $T_{\rm gas}<300$~K with the maximum value of $\alpha=1$. If the temperature restriction is lifted, the theoretical  $\sih - \log n(\HH)$ relation shifts by $-0.3$ at low densities and by $-0.8$ at high densities. 
To balance dust growth at high temperatures and to lower depletions in the diffuse ISM in models MRN3nm and EMRN3nm and to match the observational data would require an unreasonably short timescale $\tau^{\rm diff}_{\rm dest}$ of destruction in the diffuse medium in Eq~(\ref{eq:dfdtdiff}). 
We therefore conclude that the assumption of the maximum coefficient for all temperatures, commonly used in recent hydrodynamical simulation with dust \citep[e.g.,][]{McKinnon:2016ft, Bekki:2015iy, Bekki:2015hn}, tends to overestimate the dust production rates in hydrodynamic numerical simulations of galactic evolution. 

The sticking coefficients derived by  \cite{Chaabouni:2012gw} and by  \cite{LeitchDevlin:1985tp} have different dependences on the gas temperature. The latter has the maximum at about 200~K, while the former peaks at $T_{\rm gas} \sim 10$~K. The fact that both prescriptions result in the similar  $\sih - \log n(\HH)$ relations reveals that the mean depletions are not very sensitive to the choice of $\alpha$, as long as it decreases with gas temperature. CMRN3nm models with $\alpha$ from  \cite{Chaabouni:2012gw} and \cite{LeitchDevlin:1985tp} yield too high \sih{} values and a shallower slope than the values from observational data. Calculations for EMRN3nm models provide a better fit to the observations than the CMRN3nm model.

\begin{figure}
\includegraphics[width=.5\textwidth, page=1]{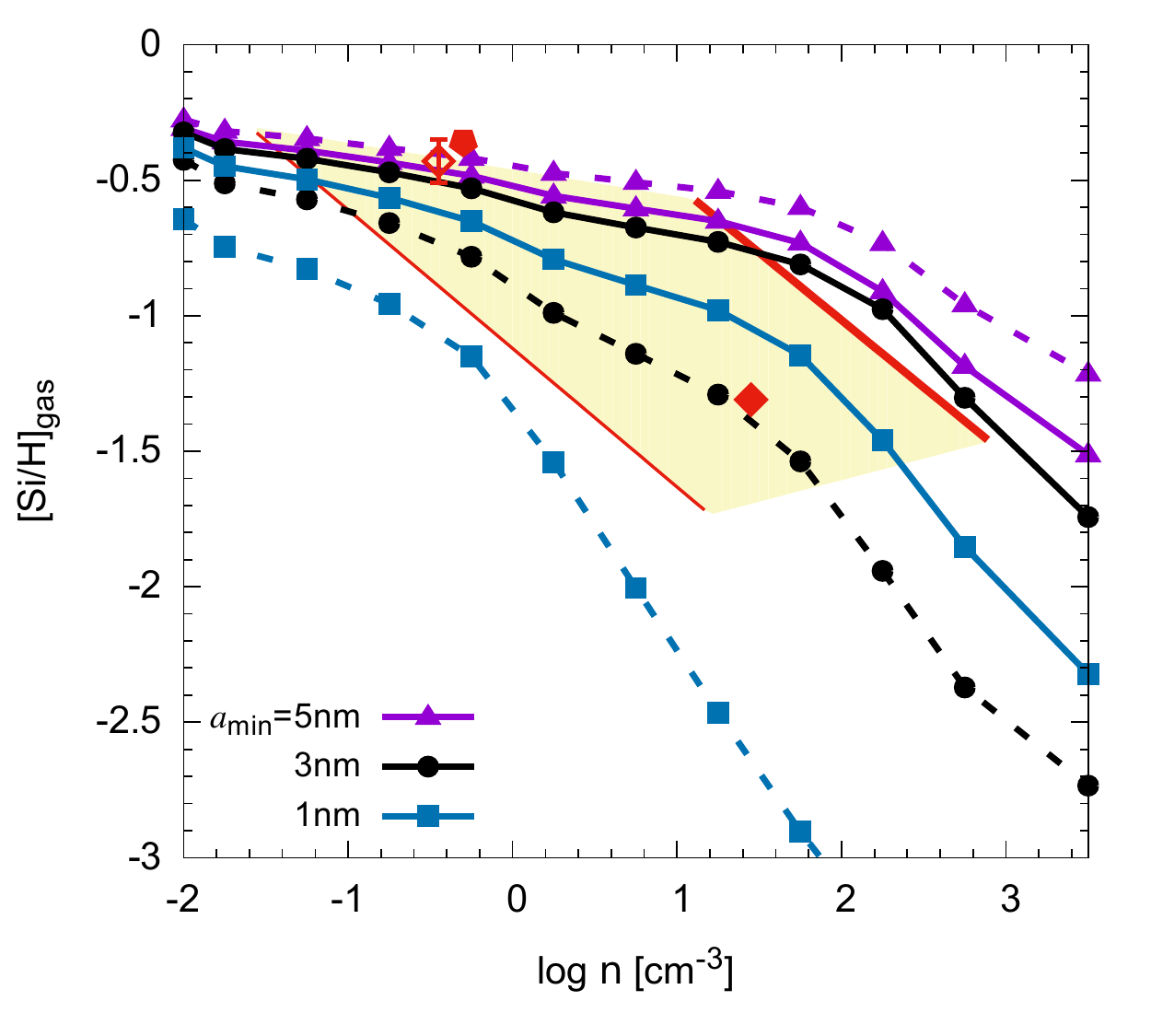}
\caption{Relation between Si depletion and gas density derived  from the final distribution of Si gas-phase abundances in the simulations for minimum grain size $a_{\min}=1$, 3, and 5~nm (lines with squares, circles, and triangles, respectively) for CMRN and ECMRN models (solid and dashed lines, respectively). The observed relation is shown in the same way as in Fig.~\ref{fig:Depl-logn}.}
\label{fig:Delpn-amin}
\end{figure}

\subsection{Dependence on the grain size distribution}\label{sec:ResMinSize}
In the previous sections we discussed various results of dust modelling for a fixed grain size distribution. The size distribution for silicate grains, in particular its lower limit, is highly uncertain parameter and varies among different models. It is determined by fitting the observed spectral energy distribution of interstellar dust in the diffuse medium \citep{Zubko:2004p4116, Siebenmorgen:2014ga}. The grain size distribution enters the rate of the dust growth in the ISM via Eqs.~(\ref{eq:dfdtgr}--\ref{eq:a3}). Variations in the grain sizes affect the final element depletions through the change in the total surface area, and additionally, through enhanced collision rates with smaller grains \citep{Weingartner:1999p6573}. 

In the following, we inspect how the variations in the lower limit of the grain size distribution affects the relation between the average \si{} depletion and gas density predicted by our models. We use the same value $a_{\max}=0.5\mu$m for the upper limit for the grain sizes, as it has has little effect on the total surface area. All other parameters are taken the same as for the reference model.  The results for models CMRN and ECMRN with $a_{\min}=1$, 3 and 5~nm for the final simulation snapshot are shown in Fig.~\ref{fig:Delpn-amin}.

The synthetic average \sih{} vs. $\log n(\HH)$ relations show double power law behaviour noted above for the reference models. With the decrease of the minimum grain size, the total surface area of grains increases and the accretion in the ISM occurs faster  resulting in steeper slopes of the  relation and  transition to the efficient growth phase at lower densities. This effect is additionally enhanced by the presence of negatively charged small grains if the collision rates  of grains with ions are included \cite[see Fig.~1 in][] {Weingartner:1999p6573}.  The differences in \si{} depletion values between ECMRN and CMRN models may reach 1~dex at high densities for $a_{\min}=1$~nm and $a_{\min}=3$~nm (Fig.~\ref{fig:Delpn-amin}). The ratio of the gas densities at which the transition to effective growth takes place roughly corresponds to the ratio of the growth timescales that for models shown in Fig.~\ref{fig:Delpn-amin} is determined by the average grain radius $\aav$ (Table~\ref{tab:avsizes}).

We do not confirm existence of the so-called Coulomb barrier for dust growth suggested recently by \cite{Ferrara:2016wp} as a process to prevent dust growth. In contrary, we find that growth by accretion in the CNM occurs even for assumption of the ``classical'' cut-off of the MRN size distribution $a_{\min}=5$~nm due to the fact that only silicate grains larger than 20~nm are positively charged. 
In this case, however, repulsion between ions and positively charged grains results in a somewhat longer accretion timescale and shallower slopes of the depletion--gas density relation than in model CMRN5nm. The mean \sih{} values in the dense gas for models CMRN5nm and ECMRN5nm are higher than the observational data (Fig.~\ref{fig:Delpn-amin}), we therefore favour the models with $a_{\min}=3$~nm. 

We conclude that silicate grains with the sizes that are smaller than the cut-off of the MRN size distribution of $a_{\min}=5$~nm are necessary to reproduce the observed slope of the $\sih - \log n(\HH)$ relation in our model. Underlying this result is the assumption that some chemical selection process prevents sticking of \si{} on very small carbonaceous grains such as desorption upon photoexcitation by UV irradiation \citep{Tielens:1998p7054, Draine:2009p6616}.

We do not take into account that the grain size distribution evolves in dense cores of molecular clouds owing to  coagulation process. \cite{Hirashita:2012dg} showed that because element depletion by accretion occurs at lower gas densities than grain coagulation, the reduction of total grain surface area by coagulation only slightly affects  the dust mass growth in the ISM. Indeed, we find that efficient dust growth occurs already in the CNM and consumes almost all gas-phase \si{} at much higher densities than the coagulation regime ($10^4\cmc$).


\section{Conclusions}\label{sec:Concl}
We develop a three-dimensional model for interstellar dust evolution as traced by \si{} abundance based on numerical hydrodynamical simulations of giant molecular clouds in a Milky-Way-like galaxy. This work focuses on the model predictions in a ring centred on the solar galactocentric radius so that they can be tested against the gas-phase \si{} abundances measured along numerous sight lines and probing various interstellar conditions in the local Milky Way. Using the depletion data and measurements of the local gas density probed by fine-structure lines of neutral carbon, we derive a relation between the average Si abundance in the gas \sih{} and the local gas density $n(\HH)$ which we use as a critical constraint for the models. 

We demonstrate that the present rate of dust re-formation in the ISM of $0.1\Msgyrpc$ in the solar neighbourhood is substantially larger than the dust injection rate by stars ($0.003\Msgyrpc$).  Most of dust growth occurs in the density range $5\cmc < n \leqslant 50\cmc$ accounting for a half of the total growth rate.  We find that the dust mass growth takes place in the cold neutral medium for all considered values of the minimum grain size $a_{\min}$ from 1 to 5 nm, despite the Coulomb repulsion between positive impinging ions and large positively charged grains. Because of negatively charged small grains, the accretion timescale for $a_{\min}\lesssim 3$~nm is shorter with account of Coulomb interactions.

The model includes destruction of grains by energetic feedback from SNe that occur mostly in the GMCs in the hydrodynamical simulations used for the dust modelling. This destruction process has a relatively long timescale of 1~Gyr and tends to over-predict the dust abundances in the WNM. The latter constraint can be only reproduced with the implementation of an additional destruction process in the diffuse ISM representing non-correlated SNe from run-away massive stars and type Ia SNe. We derive an average lifetime of silicate grains against destruction of 350~Myr for our best-fit model. Our findings re-enforce the assumption that destruction of grains in the diffuse ISM by single SNe is a dominant pathway of dust destruction. 

We calculate the probability density distributions of the \sih{} and their corresponding mean values and standard deviations for different gas densities. The resulting relation between the average \sih{} and the gas density derived for dust distribution at the end of the simulations is used to constrain the dust growth model by comparison with the observed relation. We find that the assumption of a constant sticking coefficient adopted in recent hydrodynamics models with dust significantly overestimates the dust growth rates. The observed  $\sih - \log n(\HH)$ relation requires a sticking coefficient that decreases with temperature. A simple assumption that the growth is limited to the CNM and clouds with $T<300$~K provides a slightly better fit in the CNM density range than the provisional models with temperature-dependent sticking coefficients from \cite{Chaabouni:2012gw} and \cite{LeitchDevlin:1985tp}. Including the enhanced collision rates of cations with grains in the CNM and adopting the lower limit for the grain size of $\lesssim 3$~nm in the dust growth model are crucial to reproduce the observed slope of the $\sih - \log n(\HH)$ relation. 

The dust evolution model proposed in this work does not require the presence of a population of very small silicate grains like the one known to exist for carbonaceous grains \citep{Draine:1985ei, Draine:2001p4105}. Including the 1~nm grains in the model leads to overly high \si{} depletions in the entire simulation volume. If such a population exists, then an additional mechanism, other than sputtering in SN shocks, has to be included in the model to remove accreted \si{} from the grain surfaces and provide the observed levels of \si{} depletion in the ISM. 

\acknowledgments

S.Z. acknowledges support by the Forschungsgemeinschaft through SPP 1573: “Physics of the Interstellar Medium”. We are grateful to Nikolai Voshchinnikov and Simon Glover for reading the manuscript. We thank the referee for useful comments that helped to improve the manuscript. We gratefully acknowledge the Max Planck Computing and Data Facility for providing their user support and computing time on the Odin and Hydra clusters. C.D. acknowledges support by the European Research Council under the European Community’s Seventh Framework Programme (FP7/2011-2016 grant agreement no. 280104, LOCALSTAR). The findings reported in Section 4 of this paper arose from conclusions supported by HST archival program numbers 09534 and 10279, which were provided by NASA through grants from the Space Telescope Science Institute (STScI), which is operated by the Association of Universities for Research in Astronomy, Incorporated, under NASA contract NAS5-26555.

\bibliography{/Users/lana/Documents/AllFromPapers}

\begin{thebibliography}{}
\expandafter\ifx\csname natexlab\endcsname\relax\def\natexlab#1{#1}\fi

\bibitem[{Asano {et~al.}(2013{\natexlab{a}})Asano, Takeuchi, Hirashita, \&
  Inoue}]{Asano:2013kl}
Asano, R.~S., Takeuchi, T.~T., Hirashita, H., \& Inoue, A.~K.
  2013{\natexlab{a}}, Earth, Planets and Space, 65, 213

\bibitem[{Asano {et~al.}(2013{\natexlab{b}})Asano, Takeuchi, Hirashita, \&
  Nozawa}]{Asano:2013hg}
Asano, R.~S., Takeuchi, T.~T., Hirashita, H., \& Nozawa, T. 2013{\natexlab{b}},
  Monthly Notices of the Royal Astronomical Society, 432, 637

\bibitem[{Bate {et~al.}(1995)Bate, Bonnell, \& Price}]{Bate:1995wm}
Bate, M.~R., Bonnell, I.~A., \& Price, N.~M. 1995, Monthly Notices of the Royal
  Astronomical Society, 277, 362

\bibitem[{Bekki(2013)}]{Bekki:2013iw}
Bekki, K. 2013, Monthly Notices of the Royal Astronomical Society, 436, 2254

\bibitem[{Bekki(2015{\natexlab{a}})}]{Bekki:2015iy}
---. 2015{\natexlab{a}}, The Astrophysical Journal, 799, 166

\bibitem[{Bekki(2015{\natexlab{b}})}]{Bekki:2015hn}
---. 2015{\natexlab{b}}, Monthly Notices of the Royal Astronomical Society,
  449, 1625

\bibitem[{Benz {et~al.}(1990)Benz, Cameron, Press, \& Bowers}]{Benz:1990ct}
Benz, W., Cameron, A. G.~W., Press, W.~H., \& Bowers, R.~L. 1990, The
  Astrophysical Journal, 348, 647

\bibitem[{Binney \& Tremaine(1987)}]{Binney:1987vb}
Binney, J., \& Tremaine, S. 1987, {Galactic dynamics} (Princeton: Princeton
  University Press)

\bibitem[{Bocchio {et~al.}(2014)Bocchio, Jones, \& Slavin}]{Bocchio:2014es}
Bocchio, M., Jones, A.~P., \& Slavin, J.~D. 2014, Astronomy and Astrophysics,
  570, A32

\bibitem[{Calura {et~al.}(2008)Calura, Pipino, \& Matteucci}]{Calura:2008p1752}
Calura, F., Pipino, A., \& Matteucci, F. 2008, Astronomy and Astrophysics, 479,
  669

\bibitem[{Calzetti(2013)}]{2013seg..book..419C}
Calzetti, D. 2013, in Secular Evolution of Galaxies, ed. J.~Falc{\'o}n-Barroso
  \& J.~H. Knapen (Secular Evolution of Galaxies), 419

\bibitem[{Chaabouni {et~al.}(2012)Chaabouni, Bergeron, Baouche, Dulieu, Matar,
  Congiu, Gavilan, \& Lemaire}]{Chaabouni:2012gw}
Chaabouni, H., Bergeron, H., Baouche, S., {et~al.} 2012, Astronomy and
  Astrophysics, 538, A128

\bibitem[{Christensen {et~al.}(2012)Christensen, Quinn, Governato, Stilp, Shen,
  \& Wadsley}]{Christensen:2012dw}
Christensen, C., Quinn, T., Governato, F., {et~al.} 2012, Monthly Notices of
  the Royal Astronomical Society, 425, 3058

\bibitem[{Cox \& Gomez(2002)}]{Cox:2002fz}
Cox, D.~P., \& Gomez, G.~C. 2002, THE ASTROPHYSICAL JOURNAL SUPPLEMENT SERIES,
  142, 261

\bibitem[{Dartois {et~al.}(2004)Dartois, Mu~Oz~Caro, Deboffle, \&
  D'hendecourt}]{Dartois:2004p466}
Dartois, E., Mu~Oz~Caro, G.~M., Deboffle, D., \& D'hendecourt, L. 2004,
  Astronomy and Astrophysics, 423, L33

\bibitem[{de~Bennassuti {et~al.}(2014)de~Bennassuti, Schneider, Valiante, \&
  Salvadori}]{deBennassuti:2014by}
de~Bennassuti, M., Schneider, R., Valiante, R., \& Salvadori, S. 2014, Monthly
  Notices of the Royal Astronomical Society, 445, 3039

\bibitem[{D'Hendecourt {et~al.}(1985)D'Hendecourt, Allamandola, \&
  Greenberg}]{DHendecourt:1985p867}
D'Hendecourt, L.~B., Allamandola, L.~J., \& Greenberg, J.~M. 1985, Astronomy
  and Astrophysics, 152, 130

\bibitem[{Dobbs(2008)}]{Dobbs:2008ez}
Dobbs, C.~L. 2008, Monthly Notices of the Royal Astronomical Society, 391, 844

\bibitem[{Dobbs {et~al.}(2011)Dobbs, Burkert, \& Pringle}]{Dobbs:2011gp}
Dobbs, C.~L., Burkert, A., \& Pringle, J.~E. 2011, Monthly Notices of the Royal
  Astronomical Society, 417, 1318

\bibitem[{Dobbs \& Pringle(2013)}]{Dobbs:2013hb}
Dobbs, C.~L., \& Pringle, J.~E. 2013, Monthly Notices of the Royal Astronomical
  Society, 432, 653

\bibitem[{Dorschner \& Henning(1995)}]{Dorschner:1995p7228}
Dorschner, J., \& Henning, T. 1995, The Astronomy and Astrophysics Review, 6,
  271

\bibitem[{Draine(1990)}]{Draine:1990p495}
Draine, B.~T. 1990, in Astronomical Society of the Pacific, The evolution of
  the interstelllar medium, ed. L.~Blitz, 193--205

\bibitem[{Draine(2009)}]{Draine:2009p6616}
Draine, B.~T. 2009, in ASP Conf. Ser. 414, Cosmic Dust - Near And Far, ed.
  T.~Henning, E.~Gr{\"u}n, \& A.~Steinacker, 453

\bibitem[{Draine \& Anderson(1985)}]{Draine:1985ei}
Draine, B.~T., \& Anderson, N. 1985, The Astrophysical Journal, 292, 494

\bibitem[{Draine \& Li(2001)}]{Draine:2001p4105}
Draine, B.~T., \& Li, A. 2001, The Astrophysical Journal, 551, 807

\bibitem[{Draine \& Salpeter(1979)}]{Draine:1979p1036}
Draine, B.~T., \& Salpeter, E.~E. 1979, Astrophysical Journal, 231, 438

\bibitem[{Dwek(1998)}]{Dwek:1998p67}
Dwek, E. 1998, Astrophysical Journal, 501, 643

\bibitem[{Dwek(2005)}]{Dwek:2005p1018}
Dwek, E. 2005, in AIP Conference Proceedings, ed. C.~Popescu \& R.~Tuffs (AIP),
  103--122

\bibitem[{Dwek {et~al.}(2007)Dwek, Galliano, \& Jones}]{Dwek:2007p496}
Dwek, E., Galliano, F., \& Jones, A.~P. 2007, The Astrophysical Journal, 662,
  927

\bibitem[{Dwek \& Scalo(1980)}]{Dwek:1980p490}
Dwek, E., \& Scalo, J.~M. 1980, Astrophysical Journal, 239, 193

\bibitem[{Ferrara {et~al.}(2016)Ferrara, Viti, \& Ceccarelli}]{Ferrara:2016wp}
Ferrara, A., Viti, S., \& Ceccarelli, C. 2016, 1606.07214

\bibitem[{Ferrarotti \& Gail(2006)}]{Ferrarotti:2006p993}
Ferrarotti, A.~S., \& Gail, H.-P. 2006, Astronomy and Astrophysics, 447, 553

\bibitem[{Fitzpatrick \& Massa(2007)}]{Fitzpatrick:2007p6352}
Fitzpatrick, E.~L., \& Massa, D. 2007, The Astrophysical Journal, 663, 320

\bibitem[{Forbes {et~al.}(2016)Forbes, Krumholz, Goldbaum, \&
  Dekel}]{Forbes:2016bu}
Forbes, J.~C., Krumholz, M.~R., Goldbaum, N.~J., \& Dekel, A. 2016, eprint
  arXiv:1605.00650, 1605.00650

\bibitem[{Glover \& Mac~Low(2007)}]{Glover:2007p2460}
Glover, S. C.~O., \& Mac~Low, M.-M. 2007, THE ASTROPHYSICAL JOURNAL SUPPLEMENT
  SERIES, 169, 239

\bibitem[{Grassi {et~al.}(2011)Grassi, Krstic, Merlin, Buonomo, Piovan, \&
  Chiosi}]{Grassi:2011gg}
Grassi, T., Krstic, P., Merlin, E., {et~al.} 2011, 533, A123

\bibitem[{Haris {et~al.}(2016)Haris, Parvathi, Gudennavar, Bubbly, Murthy, \&
  Sofia}]{Haris:2016bg}
Haris, U., Parvathi, V.~S., Gudennavar, S.~B., {et~al.} 2016, The Astronomical
  Journal, 151, 143

\bibitem[{Hirashita(1999)}]{Hirashita:1999p2082}
Hirashita, H. 1999, The Astrophysical Journal, 522, 220

\bibitem[{Hirashita(2012)}]{Hirashita:2012dg}
---. 2012, Monthly Notice of the Royal Astronomical Society, 422, 1263

\bibitem[{Hirashita \& Kuo(2011)}]{Hirashita:2011jr}
Hirashita, H., \& Kuo, T.-M. 2011, Monthly Notice of the Royal Astronomical
  Society, 416, 1340

\bibitem[{Hirashita {et~al.}(2015)Hirashita, Nozawa, Villaume, \&
  Srinivasan}]{Hirashita:2015et}
Hirashita, H., Nozawa, T., Villaume, A., \& Srinivasan, S. 2015, Monthly
  Notices of the Royal Astronomical Society, 454, 1620

\bibitem[{Hollenbach \& Salpeter(1971)}]{Hollenbach:1971hu}
Hollenbach, D., \& Salpeter, E.~E. 1971, The Astrophysical Journal, 163, 155

\bibitem[{Hu {et~al.}(2016)Hu, Naab, Walch, Glover, \& Clark}]{Hu:2016ek}
Hu, C.-Y., Naab, T., Walch, S., Glover, S. C.~O., \& Clark, P.~C. 2016, Monthly
  Notices of the Royal Astronomical Society, 458, 3528

\bibitem[{Jenkins(1987)}]{Jenkins:1987wz}
Jenkins, E.~B. 1987, Interstellar Processes, 134, 533

\bibitem[{Jenkins(2009)}]{Jenkins:2009p2144}
---. 2009, The Astrophysical Journal, 700, 1299

\bibitem[{Jenkins(2013)}]{Jenkins:2013kb}
---. 2013, The Astrophysical Journal, 764, 25

\bibitem[{Jenkins {et~al.}(1986)Jenkins, Savage, \& Spitzer}]{Jenkins:1986cd}
Jenkins, E.~B., Savage, B.~D., \& Spitzer, L.~J. 1986, The Astrophysical
  Journal, 301, 355

\bibitem[{Jenkins \& Tripp(2011)}]{Jenkins:2011by}
Jenkins, E.~B., \& Tripp, T.~M. 2011, The Astrophysical Journal, 734, 65

\bibitem[{Jones {et~al.}(1996)Jones, Tielens, \& Hollenbach}]{Jones:1996p6593}
Jones, A., Tielens, A., \& Hollenbach, D. 1996, The Astrophysical Journal, 469,
  740

\bibitem[{Jones {et~al.}(1994)Jones, Tielens, Hollenbach, \&
  McKee}]{Jones:1994p1037}
Jones, A.~P., Tielens, A. G. G.~M., Hollenbach, D.~J., \& McKee, C.~F. 1994,
  Astrophysical Journal, 433, 797

\bibitem[{Joseph(1988)}]{Joseph:12p7198}
Joseph, C. 1988, The Astrophysical Journal, 335, 157

\bibitem[{Joseph {et~al.}(1986)Joseph, Snow, Seab, \&
  Crutcher}]{Joseph:1986p6009}
Joseph, C.~L., Snow, T.~P., Seab, C.~G., \& Crutcher, R.~M. 1986, Astrophysical
  Journal, 309, 771

\bibitem[{Klessen \& Glover(2016)}]{Klessen:2016ik}
Klessen, R.~S., \& Glover, S. C.~O. 2016, Star Formation in Galaxy Evolution:
  Connecting Numerical Models to Reality, Saas-Fee Advanced Course, Volume 43.
  ISBN 978-3-662-47889-9. Springer-Verlag Berlin Heidelberg, 2016, p. 85, 43,
  85

\bibitem[{Krasnokutski {et~al.}(2014)Krasnokutski, Rouill{\'e}, J{\"a}ger,
  Huisken, Zhukovska, \& Henning}]{Krasnokutski:2014bi}
Krasnokutski, S.~A., Rouill{\'e}, G., J{\"a}ger, C., {et~al.} 2014, The
  Astrophysical Journal, 782, 15

\bibitem[{Leitch-Devlin \& Williams(1985)}]{LeitchDevlin:1985tp}
Leitch-Devlin, M.~A., \& Williams, D.~A. 1985, Monthly Notices of the Royal
  Astronomical Society, 213, 295

\bibitem[{Lodders(2003)}]{Lodders:2003bf}
Lodders, K. 2003, The Astrophysical Journal, 591, 1220

\bibitem[{Lodders {et~al.}(2009)Lodders, Palme, \& Gail}]{2009LanB...4B..712L}
Lodders, K., Palme, H., \& Gail, H.-P. 2009, in Solar System, ed. J.~E.
  Tr{\"u}mper (Berlin, Heidelberg: Springer Berlin Heidelberg), 712--770

\bibitem[{Mathis {et~al.}(1977)Mathis, Rumpl, \& Nordsieck}]{Mathis:1977p750}
Mathis, J.~S., Rumpl, W., \& Nordsieck, K.~H. 1977, Astrophysical Journal, 217,
  425

\bibitem[{Mattsson \& Andersen(2012)}]{Mattsson:2012kt}
Mattsson, L., \& Andersen, A.~C. 2012, Monthly Notices of the Royal
  Astronomical Society, 423, 38

\bibitem[{McKee(1989)}]{McKee:1989p1030}
McKee, C. 1989, Interstellar Dust: Proceedings of the 135th Symposium of the
  International Astronomical Union, 135, 431

\bibitem[{McKinnon {et~al.}(2016)McKinnon, Torrey, \&
  Vogelsberger}]{McKinnon:2016ft}
McKinnon, R., Torrey, P., \& Vogelsberger, M. 2016, Monthly Notices of the
  Royal Astronomical Society, 457, 3775

\bibitem[{Mihalas \& Binney(1981)}]{Mihalas:1981vw}
Mihalas, D., \& Binney, J. 1981, {Galactic astronomy: Structure and
  kinematics}, 2nd edn., Vol.~1 (San Francisco, CA, W. H. Freeman and Co.,
  1981. 608 p.)

\bibitem[{O'Donnell \& Mathis(1997)}]{ODonnell:1997p683}
O'Donnell, J.~E., \& Mathis, J.~S. 1997, Astrophysical Journal, 479, 806

\bibitem[{Oey \& Lamb(2012)}]{Oey:2012ts}
Oey, M.~S., \& Lamb, J.~B. 2012, Four Decades of Research on Massive Stars ASP
  Conference Series, 465, 431

\bibitem[{Press {et~al.}(2007)Press, Teukolsky, Vetterling, \&
  Flannery}]{Press:2007vx}
Press, W.~H., Teukolsky, S.~A., Vetterling, W.~T., \& Flannery, B.~P. 2007,
  {Numerical Recipes: The Art of Scientific Computing}, 3rd edn. (Oracle
  Corporation: Cambridge University Press)

\bibitem[{Price \& Monaghan(2007)}]{Price:2007jo}
Price, D.~J., \& Monaghan, J.~J. 2007, Monthly Notices of the Royal
  Astronomical Society, 374, 1347

\bibitem[{Reach {et~al.}(2015)Reach, Heiles, \& Bernard}]{Reach:2015gx}
Reach, W.~T., Heiles, C., \& Bernard, J.-P. 2015, The Astrophysical Journal,
  811, 118

\bibitem[{R{\'e}my-Ruyer {et~al.}(2014)R{\'e}my-Ruyer, Madden, Galliano,
  Galametz, Takeuchi, Asano, Zhukovska, Lebouteiller, Cormier, Jones, Bocchio,
  Baes, Bendo, Boquien, Boselli, DeLooze, Doublier-Pritchard, Hughes,
  Karczewski, \& Spinoglio}]{RemyRuyer:2014uu}
R{\'e}my-Ruyer, A., Madden, S.~C., Galliano, F., {et~al.} 2014, Astronomy and
  Astrophysics, 563, A31

\bibitem[{Roman-Duval {et~al.}(2014)Roman-Duval, Gordon, Meixner, Bot, Bolatto,
  Hughes, Wong, Babler, Bernard, Clayton, Fukui, Galametz, Galliano, Glover,
  Hony, Israel, Jameson, Lebouteiller, Lee, Li, Madden, Misselt, Montiel,
  Okumura, Onishi, Panuzzo, Reach, R{\'e}my-Ruyer, Robitaille, Rubio, Sauvage,
  Seale, Sewilo, Staveley-Smith, \& Zhukovska}]{RomanDuval:2014gu}
Roman-Duval, J., Gordon, K.~D., Meixner, M., {et~al.} 2014, The Astrophysical
  Journal, 797, 86

\bibitem[{Rouill{\'e} {et~al.}(2014)Rouill{\'e}, J{\"a}ger, Krasnokutski,
  Krebsz, \& Henning}]{Rouille:2014gi}
Rouill{\'e}, G., J{\"a}ger, C., Krasnokutski, S.~A., Krebsz, M., \& Henning, T.
  2014, Faraday Discussions, 168, 449

\bibitem[{Roy {et~al.}(2013)Roy, Martin, Polychroni, Bontemps, Abergel,
  Andr{\'e}, Arzoumanian, Di~Francesco, Hill, Konyves, Nguyen-Luong, Pezzuto,
  Schneider, Testi, \& White}]{Roy:2013hm}
Roy, A., Martin, P.~G., Polychroni, D., {et~al.} 2013, The Astrophysical
  Journal, 763, 55

\bibitem[{Santini {et~al.}(2014)Santini, Maiolino, Magnelli, Lutz, Lamastra,
  Li~Causi, Eales, Andreani, Berta, Buat, Cooray, Cresci, Daddi, Farrah,
  Fontana, Franceschini, Genzel, Granato, Grazian, Le~Floc{\textquoteright}h,
  Magdis, Magliocchetti, Mannucci, Menci, Nordon, Oliver, Popesso, Pozzi,
  Riguccini, Rodighiero, Rosario, Salvato, Scott, Silva, Tacconi, Viero, Wang,
  Wuyts, \& Xu}]{Santini:2014fb}
Santini, P., Maiolino, R., Magnelli, B., {et~al.} 2014, Astronomy and
  Astrophysics, 562, A30

\bibitem[{Savage \& Sembach(1996{\natexlab{a}})}]{Savage:1996p486}
Savage, B.~D., \& Sembach, K.~R. 1996{\natexlab{a}}, Annual Review of Astronomy
  and Astrophysics, 34, 279

\bibitem[{Savage \& Sembach(1996{\natexlab{b}})}]{Savage:1996p484}
---. 1996{\natexlab{b}}, Astrophysical Journal, 470, 893

\bibitem[{Seab \& Shull(1983)}]{Seab:1983gfa}
Seab, C.~G., \& Shull, J.~M. 1983, The Astrophysical Journal, 275, 652

\bibitem[{Serra D{\'\i}az-Cano \& Jones(2008)}]{SerraDiazCano:2008p588}
Serra D{\'\i}az-Cano, L., \& Jones, A.~P. 2008, Astronomy and Astrophysics,
  492, 127

\bibitem[{Siebenmorgen {et~al.}(2014)Siebenmorgen, Voshchinnikov, \&
  Bagnulo}]{Siebenmorgen:2014ga}
Siebenmorgen, R., Voshchinnikov, N.~V., \& Bagnulo, S. 2014, Astronomy and
  Astrophysics, 561, A82

\bibitem[{Slavin {et~al.}(2015)Slavin, Dwek, \& Jones}]{Slavin:2015in}
Slavin, J.~D., Dwek, E., \& Jones, A.~P. 2015, The Astrophysical Journal, 803,
  7

\bibitem[{Spitzer(1985)}]{Spitzer:1985hr}
Spitzer, L.~J. 1985, The Astrophysical Journal, 290, L21

\bibitem[{Spitzer \& Jenkins(1975)}]{Spitzer:1975ca}
Spitzer, L.~J., \& Jenkins, E.~B. 1975, Annual Review of Astronomy and
  Astrophysics, 13, 133

\bibitem[{Tielens(1998)}]{Tielens:1998p7054}
Tielens, A. G. G.~M. 1998, The Astrophysical Journal, 499, 267

\bibitem[{Voshchinnikov \& Henning(2010)}]{Voshchinnikov:2010p6003}
Voshchinnikov, N.~V., \& Henning, T. 2010, Astronomy and Astrophysics, 517, A45

\bibitem[{Walch {et~al.}(2015)Walch, Girichidis, Naab, Gatto, Glover, Wunsch,
  Klessen, Clark, Peters, Derigs, \& Baczynski}]{Walch:2015fg}
Walch, S., Girichidis, P., Naab, T., {et~al.} 2015, Monthly Notices of the
  Royal Astronomical Society, 454, 238

\bibitem[{Watson \& Salpeter(1972)}]{Watson:6p6307}
Watson, W.~D., \& Salpeter, E.~E. 1972, The Astrophysical Journal, 174, 321

\bibitem[{Weingartner \& Draine(1999)}]{Weingartner:1999p6573}
Weingartner, J., \& Draine, B. 1999, The Astrophysical Journal, 517, 292

\bibitem[{Weingartner \& Draine(2001)}]{Weingartner:2001p461}
Weingartner, J.~C., \& Draine, B.~T. 2001, THE ASTROPHYSICAL JOURNAL SUPPLEMENT
  SERIES, 134, 263

\bibitem[{Yozin \& Bekki(2014)}]{Yozin:2014tt}
Yozin, C., \& Bekki, K. 2014, Monthly Notices of the Royal Astronomical
  Society, 439, 1948

\bibitem[{Zhukovska(2008)}]{Zhukovska:2008p7215}
Zhukovska, S. 2008, PhD thesis, Ruperto-Carola-Univesity, Heidelberg

\bibitem[{Zhukovska(2014)}]{Zhukovska:2014ey}
---. 2014, Astronomy and Astrophysics, 562, A76

\bibitem[{Zhukovska \& Gail(2009)}]{Zhukovska:2009p7232}
Zhukovska, S., \& Gail, H.-P. 2009, Astronomical Society of the Pacific
  Conference Series, 414, 199

\bibitem[{Zhukovska {et~al.}(2008)Zhukovska, Gail, \&
  Trieloff}]{Zhukovska:2008bw}
Zhukovska, S., Gail, H.-P., \& Trieloff, M. 2008, Astronomy and Astrophysics,
  479, 453

\bibitem[{Zhukovska \& Henning(2013)}]{Zhukovska:2013vg}
Zhukovska, S., \& Henning, T. 2013, Astronomy and Astrophysics, 555, A99

\bibitem[{Zubko {et~al.}(2004)Zubko, Dwek, \& Arendt}]{Zubko:2004p4116}
Zubko, V., Dwek, E., \& Arendt, R.~G. 2004, THE ASTROPHYSICAL JOURNAL
  SUPPLEMENT SERIES, 152, 211

\end{thebibliography}

\end{document}